\documentclass[conference,compsoc]{IEEEtran}
\ifCLASSOPTIONcompsoc
  \usepackage[nocompress]{cite}
\else
  \usepackage{cite}
\fi
\ifCLASSINFOpdf
\else
\fi
\usepackage{xcolor}
\usepackage{algorithm}
\usepackage{algpseudocode}
\usepackage{amsmath}
\usepackage{hyperref}
\usepackage{graphicx}
\usepackage{subcaption}
\usepackage{booktabs}
\usepackage{multirow}
\usepackage{comment}
\usepackage{tabularx}
\usepackage{makecell}
\usepackage{amssymb}
\usepackage{threeparttable}

\begin{document}
%
\title{PMKPIR: Parallel Multi-Keyword PIR with Interleaved Batching for Private Boolean Retrieval}


%


\author{
	\IEEEauthorblockN{
		Rongxi Wang\IEEEauthorrefmark{1}, 
		Guanxiong Ha\IEEEauthorrefmark{1}, 
		Chunfu Jia\IEEEauthorrefmark{1}, 
		Minfen Gao\IEEEauthorrefmark{2}, and 
		Hanmiaomiao Wang\IEEEauthorrefmark{1}
	}
	\vspace{0.15cm} 
	\IEEEauthorblockA{
		\IEEEauthorrefmark{1}College of Cryptology and Cyber Science, Nankai University\\
		Email: \{wrx, 2511223\}@mail.nankai.edu.cn, \{hagx, cfjia\}@nankai.edu.cn
	}
	\vspace{0.1cm} 
	\IEEEauthorblockA{
		\IEEEauthorrefmark{2}School of Mathematical Science, Nankai University\\
		Email: gaomf@nankai.edu.cn
	}
}

\maketitle

\begin{abstract}
Keyword Private Information Retrieval (Keyword PIR) allows users to retrieve data associated with specific keywords from a database while protecting query privacy. 
However, existing Keyword PIR schemes supporting boolean retrieval either incur high computational overhead due to multiple database traversals, or fail to support universal boolean computation. 
We propose PMKPIR, a novel keyword PIR scheme supporting both parallel multi-keyword queries and universal boolean computation. 
We utilize the single instruction multiple data (SIMD) capability of BFV homomorphic encryption to propose an interleaved batching method, and design IBPIR based on this method, achieving the private matching of multiple keywords in a single database traversal. 
Furthermore, we design a boolean query processing method, extending the parallel capability of interleaved batching from keyword querying to boolean query computation, to support universal boolean computation and accelerate boolean operations. 
Building upon IBPIR, we integrate this method to further design PMKPIR. 
To the best of our knowledge, PMKPIR is the first Keyword PIR scheme to support multi-keyword batch matching within boolean queries. 
Security analysis demonstrates that, consistent with existing PIR schemes supporting boolean retrieval, PMKPIR preserves the core security guarantees of classic PIR.  
Compared with state-of-the-art schemes, PMKPIR reduces query time, communication overhead, storage overhead, and peak memory by up to 79.06$\%$, 75.00$\%$, 97.92$\%$, and 80.26$\%$, respectively.
\end{abstract}


%
\IEEEpeerreviewmaketitle

\section{Introduction}

Private Information Retrieval (PIR) \cite{ChorGKS95} enables a client to retrieve data items from a server's database while keeping the server unaware of which data items are being queried. 
From an architectural perspective, PIR schemes are primarily categorized into multi-server and single-server PIR. 
Although multi-server PIR schemes \cite{EriguchiKN22, GongHPK24} provide information-theoretic security guarantees under the non-collusion assumption, this assumption is often difficult to ensure in real-world deployments \cite{ZhaoLHMZLCZ26}. 
Therefore, single-server PIR schemes \cite{Corrigan-GibbsH22, LazzarettiP23, Menon024, MugheesR23, ZhouLTS23}, which rely on computational security assumptions to achieve query privacy, are commonly adopted in practice. 
Under the single-server architecture, existing solutions mainly include Index PIR and Keyword PIR. 
Index PIR \cite{AngelCLS18, DavidsonPC23, HenzingerHCMV23, MugheesC021, ZhouPZS24} operates under the assumption that the client already knows the exact physical location of the target data item. 
However, data items in real-world applications are typically indexed by unique identifiers that follow natural sparsity patterns, manifesting as a \{keyword, value\} structure \cite{ZhaoLHMZLCZ26}. 
Keyword PIR \cite{ChorGN98, CeliD24, 0001LPZ000D25, LiuLWR24, MahdaviK22} is oriented towards such data structures, enabling the client to directly submit keywords to retrieve data items, thus making it more applicable to sparse data scenarios \cite{ZhaoLHMZLCZ26}. 

However, existing Keyword PIR schemes \cite{MahdaviK22, PatelSY23, LiuLWR24, BienstockPSY24, GrothKL10} are mostly limited to single-keyword private retrieval and struggle to support complex queries composed of multiple conditions. 
Boolean retrieval \cite{0021593, SaltonG83} allows users to establish logical relationships among keywords via AND, OR, and NOT operations, thereby natively supporting complex multi-condition queries. 
Private boolean retrieval further requires Keyword PIR to support the aforementioned boolean queries while protecting the privacy of queried keywords \cite{SongXZZLMX26}. 
In practical scenarios, boolean retrieval is typically built on an inverted index, where each keyword can be associated with multiple data items, and each data item can contain multiple keywords, thus forming a many-to-many relationship between keywords and data items \cite{SongXZZLMX26}. 
However, conventional Keyword PIR schemes \cite{MahdaviK22, PatelSY23, LiuLWR24, BienstockPSY24, GrothKL10} typically assume a one-to-one relationship between keywords and data items. 
Consequently, they cannot directly retrieve the complete set of data items for a single keyword, let alone perform boolean combinations across multiple sets \cite{SongXZZLMX26}. 
Song et al. \cite{SongXZZLMX26} point out that utilizing conventional Keyword PIR schemes to process many-to-many relationships introduces high overhead.

To address the aforementioned issues, Song et al. \cite{SongXZZLMX26} propose BKPIR, the first Keyword PIR scheme supporting boolean retrieval. 
It utilizes multi-hot vectors to represent the associations between keywords and documents (i.e., data items), and designs homomorphic evaluation methods for boolean operations, thereby enabling efficient private boolean retrieval over many-to-many relationships. 
However, BKPIR requires processing each queried keyword individually, and the matching for each keyword necessitates a complete traversal of the entire database. 
Consequently, its computational overhead grows linearly with the number of queried keywords. 
Subsequently, Zhao et al. \cite{ZhaoLHMZLCZ26} propose MkPIR, a multi-keyword PIR scheme, and further present an amortized version to amortize query costs. 
However, MkPIR only realizes multi-keyword conjunctive queries (i.e., boolean queries composed of keywords connected by AND) and does not yet support universal boolean computation composed of AND, OR, and NOT operators. 
In summary, \textit{realizing a Keyword PIR scheme that simultaneously supports high computational efficiency and universal boolean computation remains an open challenge.} 

We propose PMKPIR, a Keyword PIR scheme supporting parallel multi-keyword queries and universal boolean computation simultaneously. 
We utilize the single instruction multiple data (SIMD) capability of BFV \cite{FanV12} to design an interleaved batching method, encoding multiple queried keywords into different slots of a single query ciphertext. 
Based on this method, we design IBPIR, enabling the server to compute the matching results of multiple keywords in parallel during a single database traversal, thereby significantly improving computational efficiency. 
To support universal boolean computation, we further design a boolean query processing method for boolean queries, extending the parallel capability of interleaved batching from keyword matching to boolean query computation, and construct PMKPIR accordingly. 
To the best of our knowledge, PMKPIR is the first Keyword PIR scheme to support multi-keyword batch matching within boolean queries. 
The source code of our prototype implementation is publicly available at \url{https://anonymous.4open.science/r/PMKPIR-DAE9/}.

Our contributions can be summarized as follows:

\begin{itemize}
	\item We design an interleaved batching method and propose IBPIR based on this method, enabling the server to compute the matching results of multiple keywords in parallel during a single database traversal. 
	IBPIR reduces the number of database traversals and the homomorphic computation overhead required for boolean queries. 
	
	\item We propose a boolean query processing method for boolean queries, extending the parallel capability of interleaved batching to boolean query computation and supporting universal boolean computation. 
	Building upon IBPIR, we integrate this method to further construct PMKPIR. 

	\item We conduct formal security proofs and detailed experimental evaluations for PMKPIR. 
	The security proofs demonstrate that PMKPIR preserves the core security guarantees of classic PIR. 
	The experimental results show that PMKPIR effectively reduces the computational overhead for private boolean retrieval. 
	Compared with state-of-the-art schemes, PMKPIR reduces query time, communication overhead, storage overhead, and peak memory by up to 79.06$\%$, 75.00$\%$, 97.92$\%$, and 80.26$\%$, respectively. 
		
\end{itemize}

\section{Preliminaries}

\subsection{Homomorphic Encryption}
\label{subsubsec:Homomorphic Encryption}

Homomorphic encryption (HE) allows computation over encrypted data, producing ciphertexts that decrypt to the correct results of the corresponding plaintext operations. 
A HE scheme that supports both addition and multiplication is known as fully homomorphic encryption (FHE). 
FHE for arbitrary computation depths is expensive, so practical systems commonly employ leveled FHE schemes with pre-limited computation depths, such as BFV \cite{FanV12}, BGV \cite{BrakerskiGV14}, and CKKS \cite{CheonKKS17}. 
We adopt the BFV scheme. 
Modern leveled FHE supports SIMD operations \cite{SmartV14}, allowing a vector of values to be encoded into multiple slots of a single ciphertext and enabling parallel execution of the same operation across all slots. 
The homomorphic operation primitives used in this paper are defined as follows. 

\begin{itemize}
	\item $\text{CtCtAdd}(ct_1,ct_2)$: It takes two ciphertexts $ct_1$ and $ct_2$ as input and outputs their slot-wise encrypted sum. 
	\item $\text{CtPtAdd}(ct,p)$: It takes a ciphertext $ct$ and a plaintext $p$ as input and outputs their slot-wise encrypted sum. 
	\item $\text{CtCtMul}(ct_1,ct_2)$: It takes two ciphertexts $ct_1$ and $ct_2$ as input and outputs their slot-wise encrypted product. 
	\item $\text{CtPtMul}(ct,p)$: It takes a ciphertext $ct$ and a plaintext $p$ as input and outputs their slot-wise encrypted product. 
	\item $\text{CtRotate}(ct,\ell)$: It takes a ciphertext $ct$ and an integer $\ell$ as input and outputs the encrypted vector cyclically rotated by $\ell$ slots. 
	\item $\text{CtNegate}(ct)$: It takes a ciphertext $ct$ as input and outputs the ciphertext of its slot-wise negation. 
\end{itemize}

\subsection{Constant-weight Code}
\label{subsubsec:Constant-weight Code}

Constant-weight codes are binary vectors of fixed Hamming weight. 
The set of constant-weight codes with length $w$ and Hamming weight $k$ is denoted as $\text{CW}(w,k) = \{y\in\{0,1\}^w:\Vert{}y\Vert{}_1=k\}$. 
The number of distinct codewords that this set can represent is $H_{\mathrm{CW}}=\binom{w}{k}$. 
Therefore, to encode a domain of size $\vert{} KW \vert{}$, the condition $\binom{w}{k}\ge \vert{} KW \vert{}$ must be satisfied. 
We adopt the constant-weight encoding function $\text{CWEncode}(a, w, k)$ consistent with the scheme of Song et al. \cite{SongXZZLMX26} to encode an integer $a \in \{0, \ldots, \vert{} KW \vert{}-1\}$ into a constant-weight codeword. 

CwPIR \cite{MahdaviK22} proposes a scheme to determine whether two constant-weight codes are equal within a HE context. 
The basic idea is to compute the inner product of two codewords and evaluate whether this inner product equals the codeword weight $k$. 
Specifically, for any two codewords $x,y \in \text{CW}(w,k)$, their inner product $u = \langle x,y \rangle$ equals $k$ if $x=y$, and falls into the range $u \in \{0, 1, \dots, k-1\}$ if $x \ne y$. 
Based on this property, an unnormalized equality polynomial is defined as $\Phi_k(u)=\prod_{i=0}^{k-1}(u-i)$. 
If the codewords do not match ($x \ne y$), $\Phi_k(u)$ evaluates to $0$; if they match ($x = y$), the inner product is $k$, yielding $\Phi_k(k) = k!$. 
To support subsequent boolean operations, this output must be normalized to $1$ for a match and $0$ for a mismatch. 
Assuming $k!$ is invertible in the plaintext ring (i.e., $\gcd(k!,t)=1$, where $t$ is the plaintext modulus), we introduce a public normalization constant $\gamma=(k!)^{-1}\pmod t$ to define the normalized equality polynomial $P_k(u)=\gamma \Phi_k(u)\pmod t$. 
Consequently, $P_k(\langle x,y\rangle)$ properly evaluates to $1$ if $x=y$, and $0$ otherwise. 
The server can either compute $P_k$ online to obtain the normalized boolean result, or compute only $\Phi_k$ online and integrate the public constant $\gamma$ into the database plaintext. 
We adopt the latter method to save one homomorphic multiplication during the online keyword matching. 

\subsection{BFV Batching Slots}

Modern leveled FHE schemes support encoding multiple plaintexts into different slots of a single FHE plaintext, enabling parallel execution of homomorphic operations across all slots of this plaintext. 
In the BFV \cite{FanV12}, let the polynomial modulus degree be $N$ and the plaintext modulus be $t$. 
When $N$ is a power of 2 and satisfies $t \equiv 1 \pmod{2N}$, a single BFV plaintext can contain $N$ batching slots, where homomorphic addition and multiplication are equivalent to slot-wise addition and multiplication on these slots. 
In the BFV batching interface of Microsoft SEAL \cite{seal40}, these $N$ slots are logically mapped into a $2 \times (N/2)$ matrix. 
Under this representation, a row rotation operation cyclically shifts the two rows of the matrix respectively, while a column rotation operation swaps the two rows. 
Denoting the slot coordinates as $(r, c)$, executing a row rotation by $\ell$ steps moves it to $(r, (c+\ell) \bmod (N/2))$. 
Consequently, the row rotation operation only cyclically changes the column coordinates of the elements within the current row, without causing any cross-row permutations. 
To concurrently process multiple queried keywords within a single query ciphertext, we divide the BFV batching slots into $B$ non-overlapping sets. 
Each set contains $w = N/B$ slots, which we define as a \textit{logical sub-block}. 
Each logical sub-block is utilized to store a queried keyword. 
Through this organizational structure, our scheme supports the parallel processing of $B$ keywords at a time within a single ciphertext. 
The specific slot indices and rotation methods are detailed in Section \ref{subsec:Interleaved Batching Method}. 

\subsection{Inverted Index and Bitmap Representation}

An inverted index is a data structure that maps keywords to their associated document sets. 
Let the document set be $D=\{d_0, d_1, \dots, d_{m-1}\}$, and the keyword set be $KW$. 
For any keyword $kw \in KW$, its inverted index records the identifiers of all documents containing $kw$, denoted as $I(kw)=\{j \mid kw \in KW_j\}$, where $KW_j$ represents the keyword set contained in the $j$-th document. 
The inverted index allows the system to directly and rapidly locate all documents relevant to a target keyword, and is therefore widely adopted in information retrieval systems. 

To facilitate boolean queries, the inverted index for each keyword can be represented as a bit vector of length $m$, where $m$ is the total number of documents in the dataset. 
For a keyword $kw$, its bit vector of length $m$ is denoted as $vm_{kw}$, where the $j$-th bit is $1$ if and only if the document $d_j$ contains $kw$, and $0$ otherwise. 
We refer to this form of representing the keyword inverted index using a bit vector as an inverted bitmap. 

\begin{figure}[t]
	\centering
	\includegraphics[width=1.0\columnwidth]{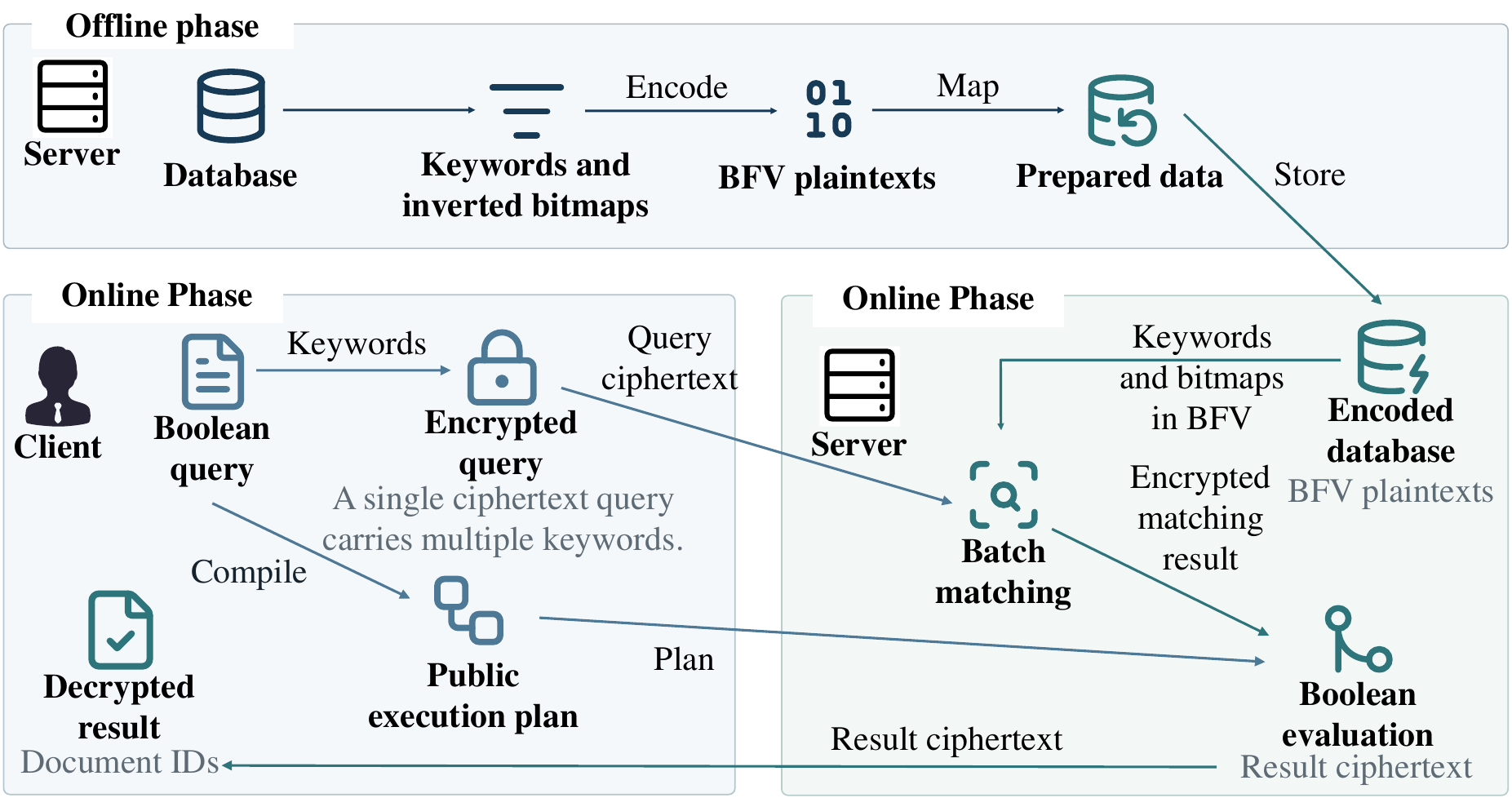}
	\vspace{-0.5cm}
	\caption{Overview of the PMKPIR workflow.}
	\label{fig:system}
	\vspace{-0.5cm}
\end{figure}

\section{System Overview}
\label{sec:overview}

As illustrated in Fig. \ref{fig:system}, the system consists of a client and a server, and is divided into an offline phase and an online phase. 
In the offline phase, the server first merges duplicate keywords in the database to extract all unique keywords. 
Subsequently, it converts the many-to-many keyword-document relationships into a linear mapping structure of keywords and their inverted bitmaps, which indicate whether specific documents contain the keyword. 
The server encodes each keyword into a constant-weight codeword, concurrently converting both the codewords and their corresponding inverted bitmaps into BFV plaintexts. 
Section \ref{subsec:details of PMKPIR} details this offline phase. 

The online phase consists of two steps: client query generation and server document retrieval. 
In the client query generation step, the client compiles the boolean query into a public execution plan and extracts the specific queried keywords (detailed in Section \ref{subsec:query compilation}). 
Then, the client sequentially encodes the processed queried keywords into constant-weight codewords and converts them into BFV plaintexts, which are then encrypted using BFV and sent to the server alongside the public execution plan. 
In the server document retrieval step, the server first completes the matching of batch queried keywords within a single database traversal, and subsequently performs boolean evaluation using these matching results. 
Specifically, the server computes the matching results between each keyword in the query ciphertext and the keywords in the database in parallel, obtaining the encrypted inverted bitmap corresponding to each queried keyword (detailed in Section \ref{subsec:details of PMKPIR}). 
Subsequently, the server merges these encrypted inverted bitmaps according to the public execution plan submitted by the client, executing the boolean evaluation at the ciphertext level to obtain the encrypted document selection bitmap (detailed in Section \ref{subsec:slot routing} and Section \ref{subsec:Homomorphic Evaluation of Boolean Queries}). 
Ultimately, the server returns the encrypted document selection bitmap to the client. 
After decrypting it, the client can output the set of document IDs that satisfy the query conditions. 
Additionally, as document payload sizes dictate the choice of document recovery schemes, the subsequent steps to extract specific document contents from the set of document IDs will be detailed in Section \ref{subsec:PMKPIR}. 

\section{Interleaved Batching and IBPIR} 
\label{sec:Keyword PIR with Batching Support}

To implement keyword PIR with batching support, we propose an interleaved batching method, which can retrieve multiple keywords simultaneously within a single database traversal. 
In this section, we detail the design of this method and the procedures of IBPIR implemented based on it. 

\subsection{Interleaved Batching Method}
\label{subsec:Interleaved Batching Method}

\begin{figure}[t]
	\centering
	\includegraphics[width=1.0\columnwidth]{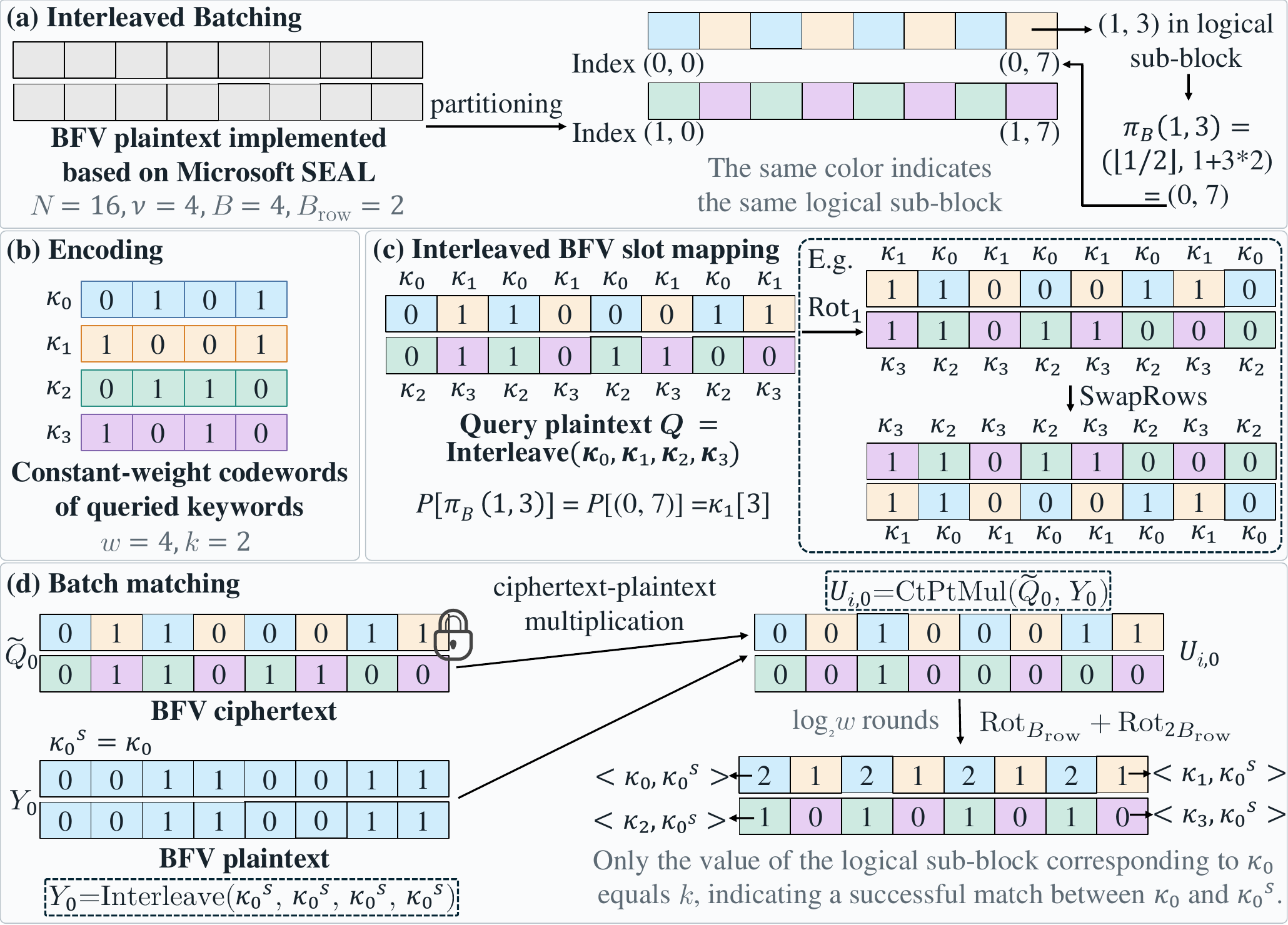}
	\vspace{-0.5cm}
	\caption{An example of the interleaved batching workflow.}
	\label{fig:Interleaved Batching}
	\vspace{-0.5cm}
\end{figure}

We design a specific BFV slot organization and batched computing method, called \textit{interleaved batching}, which supports matching multiple keywords within a single database traversal, as shown in Fig. \ref{fig:Interleaved Batching}. 
Its basic idea is to divide all batching slots within a BFV plaintext into several non-overlapping logical sub-blocks. 
Each logical sub-block is distributed non-continuously at fixed intervals within the BFV logical row to carry the constant-weight codeword of a queried keyword. 
This method enables interference-free parallel computation across different sub-blocks within the same ciphertext. 
To achieve this, the server executes rotation and accumulation operations using rotation amounts that match the fixed interval. 

As shown in Fig. \ref{fig:Interleaved Batching}(a), let the polynomial modulus degree of BFV be $N=2^\nu$, and the batching slots be arranged as a $2 \times (N/2)$ matrix according to the logical representation of Microsoft SEAL \cite{seal40}. 
If a BFV plaintext is divided into \(B=2^b\) logical sub-blocks, it can carry at most \(B\) keywords.
Each keyword occupies one logical sub-block, and corresponds to $w=N/B$ slots. 
Each row of a single BFV plaintext contains $B_{\mathrm{row}}=\frac{B}{2}$ logical sub-blocks. 
For a logical sub-block with index $q \in \{0, \dots, B-1\}$, we denote its logical row index as $r_q=\left\lfloor\frac{q}{B_{\mathrm{row}}}\right\rfloor$, and the column index of its first element within that row as $c_q=q \bmod B_{\mathrm{row}}$. 
For an element index $j \in \{0, \dots, w-1\}$ within the $q$-th logical sub-block, we define a slot mapping function $\pi_B(q,j) = \left(r_q, c_q+jB_{\mathrm{row}}\right)$, used to represent the coordinate of the $j$-th element of the $q$-th logical sub-block in the BFV logical matrix, as exemplified in Fig. \ref{fig:Interleaved Batching}(a). 
This mapping ensures that all elements across different logical sub-blocks are assigned unique coordinates, guaranteeing that their corresponding slot sets do not overlap. 

Based on the slot mapping function $\pi_B$, we define the interleaved encoding function $\text{Interleave}_B(\kappa_0,\ldots,\kappa_{B-1})$. 
This function encodes $B$ constant-weight codewords (each of length $w$) into a single BFV plaintext $P$ with $N$ slots, supporting subsequent interleaved batching. 
The generated plaintext $P$ satisfies $P[\pi_B(q,j)]=\kappa_q[j]$ for $q \in \{0, \ldots, B-1\}$ and $j \in \{0, \ldots, w-1\}$, as exemplified in Fig. \ref{fig:Interleaved Batching}(c). 
Therefore, $\text{Interleave}_B$ writes $\kappa_q$ into the slots corresponding to the $q$-th logical sub-block. 
To describe the rotation and accumulation within a logical sub-block, we use $\text{Rot}_\ell$ to denote the row rotation operation of the BFV slots. 
This operation cyclically rotates both rows of the BFV logical matrix by $\ell$ columns respectively, moving the element located at $(r,c)$ to $\left(r,\;(c+\ell)\bmod (N/2)\right)$. 
We use $\text{SwapRows}$ to denote the column rotation operation that swaps the two logical rows, moving the element located at $(r,c)$ to $(1-r,c)$. 
Both $\text{Rot}_\ell$ and $\text{SwapRows}$ are illustrated in Fig. \ref{fig:Interleaved Batching}(c) and implemented via native BFV slot permutations.

The key property of interleaved batching is that when the rotation amount is $\delta B_{\mathrm{row}}$, a row rotation does not cause elements to leave their original logical sub-blocks, where $\delta$ denotes the logical rotation step size of elements within the sub-block. 
This is because the element located at $\pi_B(q,j)$ moves to $\left(r_q,\; \bigl(c_q+(j+\delta)B_{\mathrm{row}}\bigr)\bmod (N/2)\right) = \pi_B\bigl(q,(j+\delta)\bmod w\bigr)$ after this rotation. 
Consequently, the slots before and after rotation possess the same logical sub-block index $q$, ensuring that data within different logical sub-blocks will not interfere with each other during rotation. 

In IBPIR, the keyword matching process utilizes the interleaved batching method to concurrently execute the matching process of multiple queried keywords within a single database traversal. 
As shown in Fig. \ref{fig:Interleaved Batching}(b), (c) and (d), we take a single match as an example to introduce how to achieve parallel multi-keyword matching based on this method. 
Let the constant-weight codeword of the $q$-th queried keyword be $\kappa_q$, and the constant-weight codeword of the current keyword $kw_0$ in the database be $\kappa_0^s$ (Fig. \ref{fig:Interleaved Batching}(b)). 
The client constructs the query plaintext $Q=\text{Interleave}_B(\kappa_0,\ldots,\kappa_{B-1})$, and encrypts it using BFV to obtain $\tilde{Q}$, and the server constructs the keyword BFV plaintext $Y_0=\text{Interleave}_B(\kappa_0^s,\ldots,\kappa_0^s)$ in the database. 
Fig. \ref{fig:Interleaved Batching}(c) and (d) illustrate these resulting data layouts.
After executing the element-wise ciphertext-plaintext multiplication, the $j$-th slot of the $q$-th logical sub-block contains $\kappa_q[j]\kappa_0^s[j]$. 
Subsequently, the server sequentially uses $B_{\mathrm{row}}, 2B_{\mathrm{row}}, 4B_{\mathrm{row}}, \ldots, \frac{w}{2}B_{\mathrm{row}}$ as rotation amounts to execute rotations and additions. 
Since $w$ is an integer power of two, after $\log_2w$ rounds of computation, every slot in the $q$-th logical sub-block contains $\sum_{j=0}^{w-1}\kappa_q[j]\kappa_0^s[j] = \langle \kappa_q,\kappa_0^s\rangle$. 
An inner product equal to the Hamming weight $k$ indicates a successful match between the queried keyword and the keyword in the database.
Through the subsequent constant-weight equality test (see Section \ref{subsubsec:Constant-weight Code} for details), the server verifies whether the two keywords are equal. 
Because different logical sub-blocks remain isolated, the equality tests within each sub-block operate independently, ensuring that the $B$ distinct queried keywords multiplexed in the same ciphertext do not interfere during the matching process. 

\subsection{Constructions of IBPIR}
\label{subsec:details of PMKPIR}

\textbf{Setup.} 
This phase is executed by the server to convert the $n \times m$ many-to-many keyword-document relationships into a linear structure of $n$ rows of keyword and inverted bitmap pairs, where the number of keywords is $n$ and the number of documents is $m$. 
Let $KW$ denote the public keyword set, which contains not only the existing keywords in the database but also all potential keywords permitted for querying by the protocol. 
To assign a unique constant-weight codeword to each keyword in the set, we select constant-weight encoding parameters satisfying $\binom wk\ge \vert{} KW \vert{}$. 
Additionally, we adopt a public mapping function $\text{ID}:KW \rightarrow \left\{0,\ldots,\binom wk-1\right\}$ that is a collision-free injection, used to map keywords to integers suitable for constant-weight encoding.  
Note that identical keywords in the database need to be merged into the same keyword prior to the setup phase. 
Otherwise, identically encoded keywords are simultaneously selected by the same query, and the addition of their bitmaps may produce slot values greater than one, destroying the correctness of subsequent boolean operations.

\begin{algorithm}
	\caption{Setup}
	\label{alg:setup_PMKPIR}
	\begin{algorithmic}[1]
		\renewcommand{\algorithmicrequire}{\textbf{Input:}}
		\renewcommand{\algorithmicensure}{\textbf{Output:}}
		\Require $DB$, $w$, $k$, $B$, $\gamma = (k!)^{-1} \bmod t$
		\Ensure $(Y, \widehat{M}) \in DB_{PMKPIR}$
		
		\State Choose a public keyword set $KW$ and a public mapping function $\text{ID}:KW \rightarrow \left\{0,\ldots,\binom wk-1\right\}$
		\State $f_B \gets \lceil m / w \rceil$
		\State Initialize array $Y$ of size $n$, and matrix $\widehat{M}$ of size $n \times f_B$
		\State Let $\{kw_0^s, \dots, kw_{n-1}^s\}$ be the $n$ unique keywords in $DB$
		\For{$i = 0, \dots, n - 1$}
		\For{$j = 0, \dots, m - 1$}
		\If{$kw_i^s \in KW_j$}
		\State $vm_i[j] \gets 1$
		\Else
		\State $vm_i[j] \gets 0$
		\EndIf
		\EndFor
		\State $\kappa_i^s \gets \text{CWEncode}(\text{ID}(kw_i^s))$
		\State $Y[i] \gets \text{Interleave}_B(\kappa_i^s, \dots, \kappa_i^s)$
		
		\State Split $vm_i$ into $vm_{i,0} \parallel \dots \parallel vm_{i,f_B-1}$, each of length $w$
		\For{$h = 0, \dots, f_B - 1$}
		\State $M_{i,h} \gets \text{Interleave}_B(vm_{i,h}, \dots, vm_{i,h})$
		\State $\widehat{M}[i][h] \gets \gamma \cdot M_{i,h}$
		\EndFor
		\EndFor
		\State \textbf{return} $(Y, \widehat{M})$
	\end{algorithmic}
\end{algorithm}

Algorithm \ref{alg:setup_PMKPIR} details the setup procedure. 
The server constructs an inverted bitmap $vm_i$ of length $m$ for each keyword $kw_i^s$. 
The element $vm_i[j] = 1$ if $kw_i^s \in KW_j$, and $0$ otherwise, where $KW_j$ represents the keyword set contained in the $j$-th document. 
Next, the server executes the $\text{CWEncode}$ function (Line 13) to map $kw_i^s$ into the constant-weight codeword $\kappa_i^s$. 
To align with the layout of the query ciphertext for parallel matching, the server replicates $\kappa_i^s$ $B$ times. 
The server calls $\text{Interleave}_B(\kappa_i^s,\ldots,\kappa_i^s)$ to pack these replicas into the $B$ logical sub-blocks of a BFV plaintext, obtaining the keyword plaintext $Y[i]$ (Line 14). 
Fig. \ref{fig:Interleaved Batching}(d) illustrates this resulting data layout. 
According to the interleaved batching method, a single logical sub-block has length $w$, accommodating the positional information of at most $w$ documents. 
If the total number of documents $m > w$, the bitmap $vm_i$ is partitioned into $f_B = \lceil m/w \rceil = \lceil mB/N \rceil$ chunks (Line 15), denoted as $vm_{i,0} \parallel \dots \parallel vm_{i,f_B-1}$. 
To ensure correctness in the response phase, the $B$ logical sub-blocks within a single response plaintext must store identical document positional information. 
Therefore, for each chunk index $h \in \{0, \dots, f_B-1\}$ (Line 16), the server calls $\text{Interleave}_B(vm_{i,h}, \dots, vm_{i,h})$ to replicate the chunk $vm_{i,h}$ across all $B$ sub-blocks, generating $M_{i,h}$ (Line 17). 
Finally, the scaling operation $\widehat{M}[i][h] \gets \gamma \cdot M_{i,h}$ (Line 18) is computed directly within the plaintext ring using the public normalization constant $\gamma = (k!)^{-1} \pmod t$. 
As described in Section \ref{subsubsec:Constant-weight Code}, this cancels out the extra coefficient $k!$ generated by the unnormalized equality product upon a successful keyword match, reducing the online homomorphic computational overhead.

\noindent \textit{Remark.} 
It should be noted that the setup phase only changes the plaintext representation of the data, without changing its logical structure. 
Specifically, each keyword in the database is still encoded individually, and its constant-weight codeword and all bitmap chunks are respectively written into independent plaintexts. 

\begin{algorithm}
	\caption{QueryGen}
	\label{alg:querygen}
	\begin{algorithmic}[1]
		\renewcommand{\algorithmicrequire}{\textbf{Input:}}
		\renewcommand{\algorithmicensure}{\textbf{Output:}}
		\Require Keyword list $KL = (kw_0^c, \dots, kw_{s-1}^c)$, $w$, $k$, $B$
		\Ensure $\{\widetilde{Q}_0, \dots, \widetilde{Q}_{\alpha-1}\}$
		
		\State $\alpha \gets \lceil s / B \rceil$
		\For{$\tau = 0, \dots, \alpha - 1$}
		\For{$q = 0, \dots, B - 1$}
		\If{$\tau B + q < s$}
		\State $\kappa_{\tau,q} \gets \text{CWEncode}(\text{ID}(kw_{\tau B+q}^c))$
		\Else
		\State $\kappa_{\tau,q} \gets 0^w$
		\EndIf
		\EndFor
		\State $Q_\tau \gets \text{Interleave}_B(\kappa_{\tau,0}, \dots, \kappa_{\tau,B-1})$
		\State $\widetilde{Q}_\tau \gets \text{Encrypt}_{pk}(Q_\tau)$
		\EndFor
	\end{algorithmic}
\end{algorithm}

\textbf{QueryGen.}
This phase executes on the client side. 
It uses the same constant-weight encoding parameters and public mapping function as the setup phase to batch-encode the queried keyword list into BFV plaintexts, and encrypts them using BFV before sending them to the server. 
Algorithm \ref{alg:querygen} details the QueryGen procedure. 
Given a queried keyword list $KL=(kw_0^c,\ldots,kw_{s-1}^c)$ of size $s$, the client computes the required number of query ciphertexts $\alpha=\lceil s/B \rceil$. 
To construct the $\tau$-th query plaintext $Q_\tau$ ($\tau \in \{0, \dots, \alpha-1\}$), the client sequentially prepares $B$ codewords (indexed as $q=0, \dots, B-1$). 
If the keyword index satisfies $\tau B+q < s$, the client encodes the corresponding keyword into a constant-weight codeword $\kappa_{\tau,q} = \text{CWEncode}(\text{ID}(kw_{\tau B+q}^c))$ (Line 5). 
Otherwise, it generates an all-zero codeword $\kappa_{\tau,q} = 0^w$ for padding (Line 7). 
As exemplified in Fig. \ref{fig:Interleaved Batching}(c), we call the function $\text{Interleave}_B(\kappa_{\tau,0}, \dots, \kappa_{\tau,B-1})$ to convert these $B$ constant-weight codewords into the query plaintext $Q_\tau$ (Line 10). 
Finally, the client encrypts $Q_\tau$ into $\widetilde{Q}_\tau=\text{Encrypt}_{pk}(Q_\tau)$ using the public key $pk$ (Line 11), and sends the resulting ciphertexts $\{\widetilde{Q}_0, \dots, \widetilde{Q}_{\alpha-1}\}$ to the server, where $\text{Encrypt}_{pk}(\cdot)$ denotes the BFV encryption with public key $pk$. 

\textbf{Response.}
This phase is executed by the server. 
Its goal is to use our designed interleaved batching method to parallelize the matching between the queried keywords and the database keywords within a single database traversal. 
This process generates the encrypted bitmaps that record the document IDs for each queried keyword. 
Algorithm \ref{alg:response} presents the specific steps of this phase. 
According to the query generation rules of the client, each query ciphertext $\widetilde{Q}_\tau$ contains $B$ logical sub-blocks, which respectively store the codewords $\kappa_{\tau,q}$ of different queried keywords (where $q = 0, \dots, B-1$). 
Taking $\widetilde{Q}_0$ in Fig. \ref{fig:Interleaved Batching}(d) as an example, it records the encrypted information of the queried codewords $\kappa_0, \dots, \kappa_{B-1}$. 
During the matching process, the $i$-th keyword BFV plaintext $Y_i$ in the database records the constant-weight codeword $\kappa_i^s$ corresponding to the database keyword $kw_i^s$. 
The server calculates the element-wise ciphertext-plaintext product $\widetilde{U_{i,\tau}} \gets \text{CtPtMul}(\widetilde{Q}_\tau, Y_i)$. 
After this multiplication, the $j$-th slot of the $q$-th logical sub-block in $\widetilde{U_{i,\tau}}$ stores the bitwise product of the respective codeword elements, i.e., $\kappa_{\tau,q}[j] \cdot \kappa_i^s[j]$. 
To compute the codeword inner product of the queried keyword and the keyword in the database within each logical sub-block, the server needs to accumulate the values of all $w$ slots in the sub-block. 
To achieve this, the server executes rotation and accumulation operations $\widetilde{U_{i,\tau}} \gets \widetilde{U_{i,\tau}} + \text{Rot}_{\delta B_{\text{row}}}(\widetilde{U_{i,\tau}})$, where $\delta B_{\text{row}}$ denotes the rotation step size. 
Letting $\delta = 1, 2, 4, \dots, w/2$ respectively, the server executes $\log_2 w$ rounds of these operations. 
After $\log_2 w$ rounds of calculation, the accumulations within each logical sub-block are completed synchronously, and all slots of the $q$-th logical sub-block in this ciphertext contain the inner product result $\langle \kappa_{\tau,q}, \kappa_i^s \rangle$ of the corresponding query codeword $\kappa_{\tau,q}$ and the current keyword codeword $\kappa_i^s$ in the database. 
For example, in the 0-th row of the bottom right matrix in Fig. \ref{fig:Interleaved Batching}(d), the slots at positions 0, 2, 4, and 6 record the inner product of $\kappa_0$ and $\kappa_0^s$.

\begin{algorithm}
	\caption{Response}
	\label{alg:response}
	\begin{algorithmic}[1]
		\renewcommand{\algorithmicrequire}{\textbf{Input:}}
		\renewcommand{\algorithmicensure}{\textbf{Output:}}
		\Require $k, w, B_{\text{row}}, \{\widetilde{Q}_0, \dots, \widetilde{Q}_{\alpha-1}\},$
		\Statex \hspace{1.2em} $\{(Y_i, \widehat{M}_{i,0}, \dots, \widehat{M}_{i, f_B-1})\}_{i=0}^{n-1}$
		\Ensure $\{\widetilde{R}_{\tau,h}\}_{\tau \in [0, \alpha-1], h \in [0, f_B-1]}$
		
		\For{$\tau = 0, \dots, \alpha - 1$}
		\For{$h = 0, \dots, f_B - 1$}
		\State $\widetilde{A}_{\tau,h} \gets 0$
		\EndFor
		\For{$i = 0, \dots, n - 1$}
		\State $\widetilde{U_{i,\tau}} \gets \text{CtPtMul}(\widetilde{Q}_\tau, Y_i)$
		\For{$\delta = 1, 2, 4, \dots, w/2$}
		\State $\widetilde{U_{i,\tau}} \gets \widetilde{U_{i,\tau}} + \text{Rot}_{\delta B_{\text{row}}}(\widetilde{U_{i,\tau}})$
		\EndFor
		\State $\widetilde{E}_{i,\tau} \gets \text{EvalTree}(\Phi_k, \widetilde{U_{i,\tau}})$
		\For{$h = 0, \dots, f_B - 1$}
		\State $\widetilde{A}_{\tau,h} \gets \widetilde{A}_{\tau,h} + \text{CtPtMul}(\widetilde{E}_{i,\tau}, \widehat{M}_{i,h})$
		\EndFor
		\EndFor
		\If{$k \ge 2$}
		\For{$h = 0, \dots, f_B - 1$}
		\State $\widetilde{R}_{\tau,h} \gets \text{Relinearize}(\widetilde{A}_{\tau,h})$
		\EndFor
		\Else
		\For{$h = 0, \dots, f_B - 1$}
		\State $\widetilde{R}_{\tau,h} \gets \widetilde{A}_{\tau,h}$
		\EndFor
		\EndIf
		\EndFor
	\end{algorithmic}
\end{algorithm}

Based on this inner product result, the server calls the $\text{EvalTree}$ function (Line 10) to compute the unnormalized equality polynomial $\widetilde{E}_{i,\tau} = \Phi_k(\widetilde{U_{i,\tau}})$ via a divide-and-conquer approach \cite{SongXZZLMX26}, where $\Phi_k(x) = \prod_{i=0}^{k-1}(x-i) \pmod t$. 
Subsequently, the server multiplies $\widetilde{E}_{i,\tau}$ with the corresponding bitmap chunk $\widehat{M}_{i,h}$ calculated in the setup phase, and accumulates it into the aggregated ciphertext $\widetilde{A}_{\tau,h}$, which is $\widetilde{A}_{\tau,h} = \sum_{i=0}^{n-1} \text{CtPtMul}(\widetilde{E}_{i,\tau}, \widehat{M}_{i,h})$. 
After completing the traversal of all $n$ keywords in the database and the accumulation of the corresponding bitmap chunks, the server obtains the aggregated ciphertexts $\widetilde{A}_{\tau,h}$. 
Due to interleaved batching, the aggregated ciphertext $\widetilde{A}_{\tau,h}$ packs the encrypted bitmap chunks for $B$ queried keywords within its logical sub-blocks. 
The server then derives the result ciphertext $\widetilde{R}_{\tau,h}$ from $\widetilde{A}_{\tau,h}$. 
The $\text{Relinearize}$ function (Line 17) is used to perform homomorphic relinearization, compressing the expanded ciphertext generated by homomorphic multiplication back to its initial dimension. 

Finally, the server returns the result ciphertexts to the client. 
The client decrypts and concatenates them to recover the complete document selection bitmap, where the positions with a value of $1$ correspond to the document IDs satisfying the query conditions. 
The subsequent retrieval of document content belongs to an independent payload recovery phase. 
Depending on the document size $S_\ell$ and the backend encoding capacity $P_{cap}$, the system adopts Multi-query PIRANA \cite{LiuLWR24} to retrieve multiple documents in a batch when $S_\ell \le P_{cap}$, or SpiralStreamPack \cite{MenonW22} for chunked streaming transmission when $S_\ell > P_{cap}$. 

\section{Boolean Query Processing and PMKPIR}

To support universal boolean computation, we propose a boolean query processing method, which extends the parallel capability of interleaved batching to boolean query computation. 
Specifically, this method consists of three steps: query compilation, slot routing, and boolean evaluation. 
Query compilation (Section \ref{subsec:query compilation}) is executed by the client during the \textbf{QueryGen} phase to transform the syntax tree into a public execution plan. 
During the \textbf{Response} phase, after completing the batch keyword matching, the server aligns the encrypted inverted bitmaps corresponding to the queried keywords via slot routing (Section \ref{subsec:slot routing}). 
Subsequently, it homomorphically evaluates the boolean queries (Section \ref{subsec:Homomorphic Evaluation of Boolean Queries}) to yield the final document selection bitmap. 
In this section, we detail the design of this method and the procedures of PMKPIR implemented based on it. 

\subsection{Query Compilation}
\label{subsec:query compilation}

The client needs to compile the boolean query into a public execution plan to guide the server in completing the boolean operations using homomorphic computation. 
We propose a hybrid compilation strategy combining bounded local DNFs with syntax tree-based evaluation. 
This strategy utilizes the parallelism of interleaved batching while avoiding the exponential scale explosion caused by expanding the entire query into Disjunctive Normal Form (DNF). 

\begin{figure}[t]
	\centering
	\includegraphics[width=1.0\columnwidth]{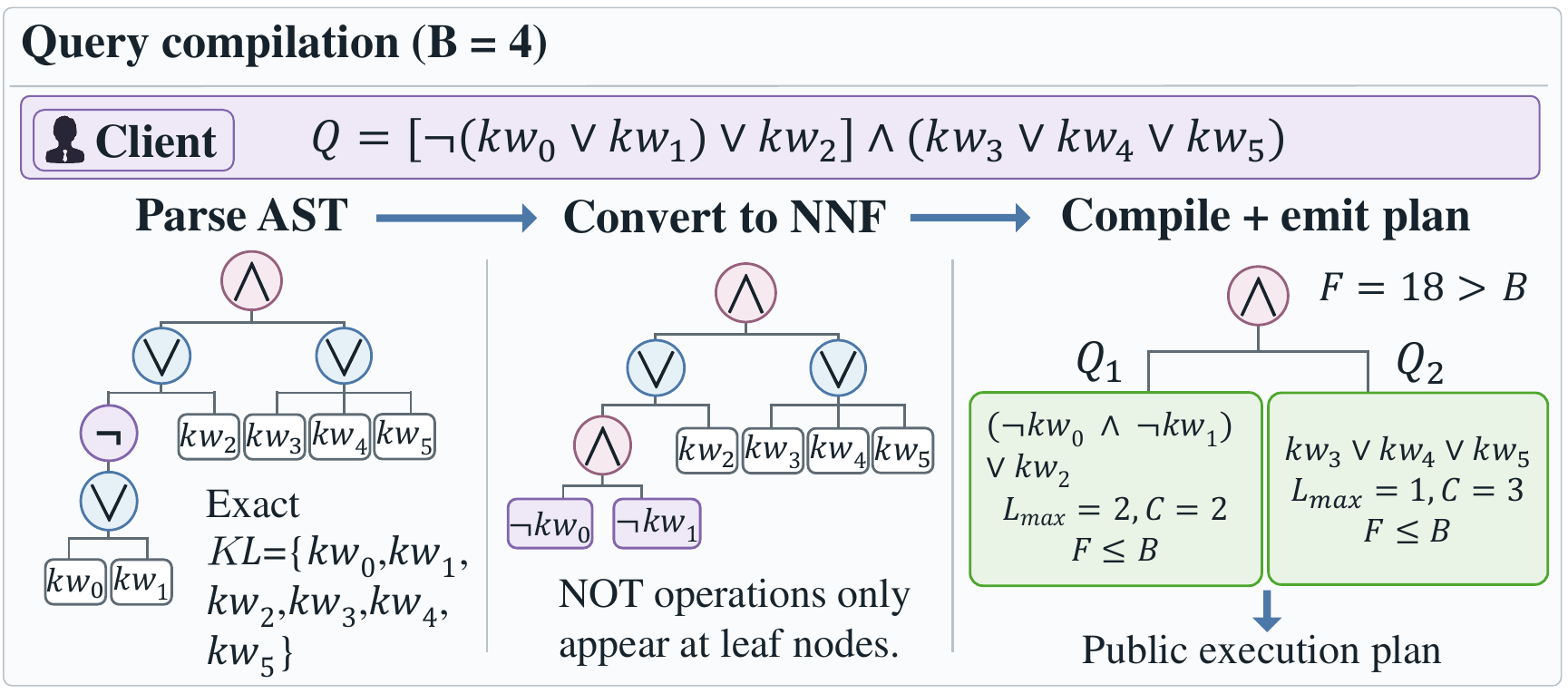}
	\vspace{-0.4cm}
	\caption{An example of the query compilation workflow.}
	\label{fig:compilation}
	\vspace{-0.7cm}
\end{figure}

As illustrated in Fig. \ref{fig:compilation}, the client's processing of the boolean query is divided into two steps: keyword extraction and syntax tree conversion. 
During the QueryGen phase (Section \ref{subsec:details of PMKPIR}), the client parses the boolean query to extract all contained keywords, forming the queried keyword list $KL=(kw_0^c,\ldots,kw_{s-1}^c)$. 
Simultaneously, the client parses the boolean query into an abstract syntax tree (AST). 
As shown in Fig. \ref{fig:compilation}, it pushes the NOT operators in the syntax tree down to the leaf nodes via De Morgan's laws, converting the expression into Negation Normal Form (NNF). 
After this conversion, the internal nodes of the syntax tree retain only AND and OR operators, while the leaf nodes correspond to the matching results of specific keywords in the keyword list $KL$. 
This separation of data (the queried keyword list $KL$) and logical structure (the syntax tree) enables the server to independently complete batch keyword matching in the ciphertext domain.  
Subsequently, based on the negation flags in the public execution plan, the server directly performs homomorphic negations (i.e., $x \rightarrow 1-x$) on the corresponding encrypted inverted bitmaps, feeding these results into subsequent homomorphic computations. 

After completing the NNF conversion, the client traverses each subtree in the syntax tree bottom-up, 
attempting to convert it into a local DNF for parallel evaluation across the logical sub-blocks of a ciphertext.
A single query ciphertext contains only $B$ logical sub-blocks. 
To evaluate the required capacity without unrolling the syntax tree and avoid ciphertext overflow from blind expansion, the client precalculates the subtree's actual maximum clause length $L_{\max}$ and clause count $C$ as if fully converted to a DNF. 
Specifically: 
(1) a single leaf node (i.e., an isolated keyword or its negation) is the simplest clause, thus $C=1$ and $L_{\max}=1$; 
(2) encountering an OR operation introduces parallel logical branches; 
thus, the total number of clauses $C$ is the sum of the clause counts of its child nodes ($C = \sum C_i$), and the maximum clause length $L_{\max}$ is the maximum among them ($L_{\max} = \max_i \{L_{\max,i}\}$), where $C_i$ and $L_{\max,i}$ denote the clause count and maximum clause length of the $i$-th child node, respectively; 
(3) encountering an AND operation means the clauses of all child nodes need to be combined via Cartesian product, causing the clause count to be multiplied ($C = \prod C_i$), and since the newly combined clauses contain the keywords from all original branches, their lengths are summed ($L_{\max} = \sum L_{\max, i}$). 
As shown in Fig. \ref{fig:compilation}, taking the subtree $\neg(kw_0 \lor kw_1) \lor kw_2$ as an example, it is first converted into $(\neg kw_0 \land \neg kw_1) \lor kw_2$. 
For the bottom AND node, a Cartesian product on its leaf nodes yields its clause count $C_{\text{AND}}=1 \times 1=1$ and maximum length $L_{\max,\text{AND}}=1+1=2$. 
This result is then merged with the outer leaf $kw_2$ via the OR node to yield the parameters for the entire subtree as $C_{\text{OR}}=C_{\text{AND}}+1=2$ and $L_{\max,\text{OR}}=\max(L_{\max,\text{AND}}, 1)=2$, where the subscripts denote the corresponding logical nodes. 

The client calculates $F = L_{\max} \cdot C$ to evaluate the total number of logical sub-blocks required for this local DNF expansion. 
If $F \le B$, the client accepts this local DNF expansion to utilize interleaved batching for efficient parallel evaluation. 
If $F > B$, it indicates that the expanded logical structure exceeds the capacity limit of a single ciphertext. 
The client stops the current expansion and instead recursively processes its child nodes. 
For the internal nodes where expansion is stopped, the client retains their original AND or OR operators in the public execution plan, instructing the server to subsequently take the encrypted bitmaps returned by each child node directly as operands and execute homomorphic operations in topological order. 
This fallback mechanism avoids the exponential size explosion caused by expanding the query into DNF, ensuring that correct computation results can be obtained for arbitrarily complex boolean syntax trees. 

Finally, the client completes the traversal of the entire syntax tree, integrates all accepted local DNF structures with the original logical operations retained by the fallback mechanism, serializes them according to their topological dependencies, and generates the public execution plan. 
This plan dictates the execution order for the server to perform homomorphic negations, parallel evaluations of local DNFs, and homomorphic intersections and unions within the ciphertext domain. 
The server performs homomorphic computation based on this public execution plan to realize boolean operations over the ciphertexts. 
Additionally, when the number of documents $m > w$, the document bitmap is partitioned into $f_B=\lceil m/w\rceil$ chunks. 
The generated public execution plan is universal across all chunks; during actual evaluation, the server repeatedly executes the same homomorphic operations on each chunk. 

\subsection{Slot Routing}
\label{subsec:slot routing}

\begin{figure}[t]
	\centering
	\includegraphics[width=1.0\columnwidth]{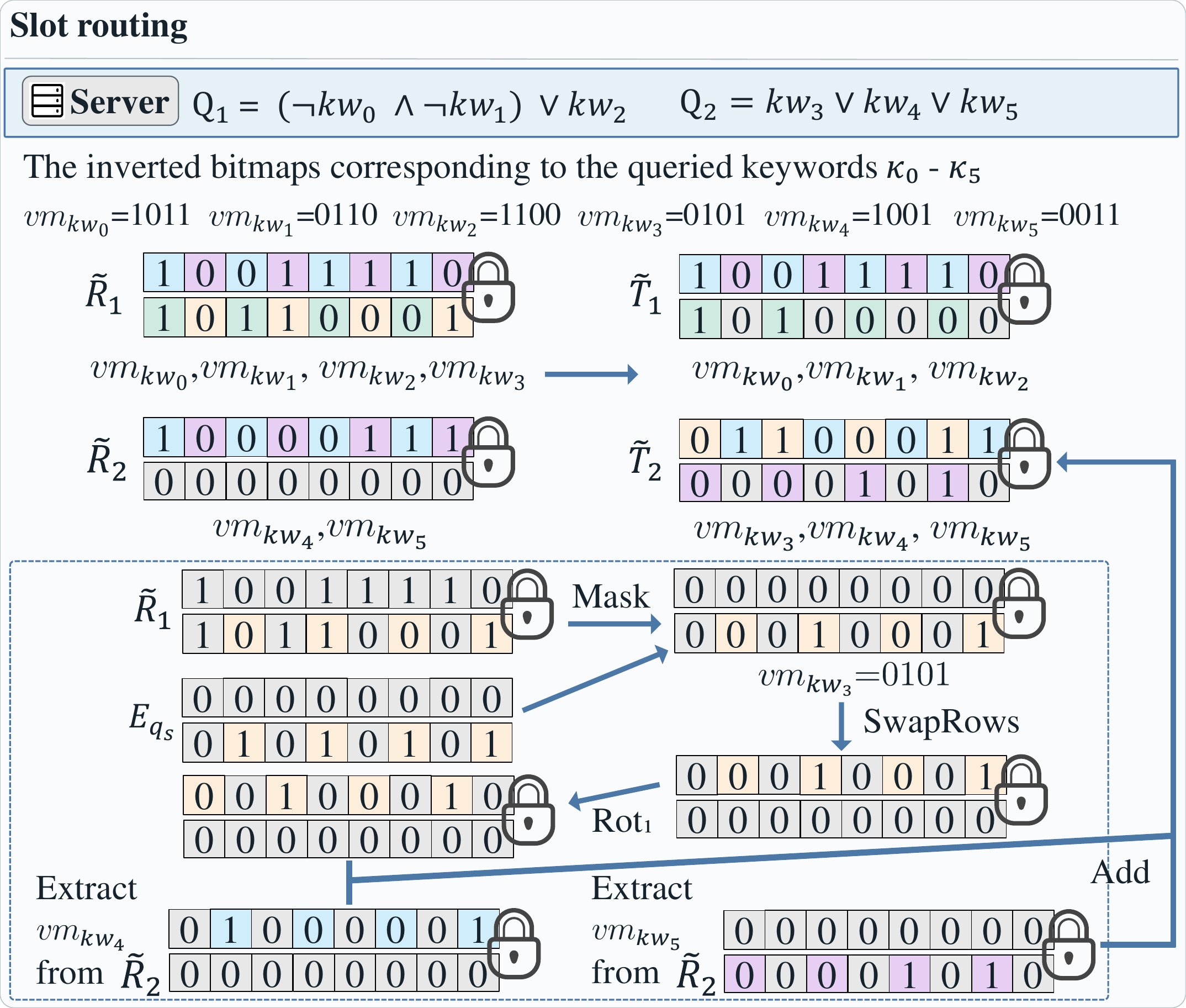}
	\vspace{-0.5cm}
	\caption{An example of the slot routing workflow.}
	\label{fig:routing}
	\vspace{-0.6cm}
\end{figure}

During the Response phase, the server executes ciphertext slot routing according to the public execution plan to aggregate all operand bitmaps of a local DNF into a single ciphertext. 
The parallel computation of the local DNF requires all its operand bitmaps to be located within the same ciphertext. 
However, the bitmaps output by multi-keyword matching are usually dispersed across different ciphertexts (as shown in Fig. \ref{fig:routing}, the bitmaps required by $Q_2$ are distributed in $\tilde{R}_1$ and $\tilde{R}_2$). 
Following the multi-keyword matching in the Response phase, the server executes ciphertext slot routing according to the public execution plan, aggregating all bitmaps of the same local DNF into different logical sub-blocks within the same ciphertext. 

To organize all inputs of the same local DNF into a unified ciphertext, the server routes each encrypted bitmap from its source logical sub-block $q_s$ in the source ciphertext to the target logical sub-block $q_d$ in the target ciphertext (e.g., routing $vm_{kw_3}$ from $\tilde{R}_1$ to $\tilde{T}_2$ in Fig. \ref{fig:routing}). 
This process sequentially includes four steps: mask extraction, row alignment, column rotation, and homomorphic addition. 
In the mask extraction step, the server defines the extraction mask $E_{q_s}$, such that its value is 1 at the position of the source sub-block $q_s$ and 0 at the remaining positions. 
Through the element-wise ciphertext-plaintext multiplication $\text{CtPtMul}(\tilde{R}, E_{q_s})$, the server extracts the source logical sub-block data, where $\tilde{R}$ is the encrypted bitmap to be routed. 
As shown in Fig. \ref{fig:routing}, to extract the bitmap $vm_{kw_3}$, the server generates the extraction mask $E_{q_s}$ corresponding to its sub-block position and computes $\text{CtPtMul}(\tilde{R}_1, E_{q_s})$ to isolate $vm_{kw_3}$ from the ciphertext. 
In the row alignment step, because the source sub-block $q_s$ and the target sub-block $q_d$ may map to different logical rows (i.e., $r_{q_s} \ne r_{q_d}$), the server calls the $\text{SwapRows}$ function to swap the upper and lower rows of the ciphertext matrix. 
In the column rotation step, the server calculates the column offset $\Delta_c = (c_{q_s} - c_{q_d}) \bmod (N/2)$. 
Here, $c_{q_s}$ and $c_{q_d}$ denote the column indices of the first elements in the source and target sub-blocks, respectively. 
Based on this offset, the server calls the column rotation function $\text{Rot}_{\Delta_c}$ with a step size of $\Delta_c$ to move the data to the target logical sub-block $q_d$ of the target ciphertext $\tilde{T}$. 
Taking the routing of \(vm_{kw_3}\) in Fig. \ref{fig:routing} as an example, the server computes \(Z=\text{CtPtMul}(\tilde{R}_1,E_{q_s})\) to extract \(vm_{kw_3}\) from its source sub-block. 
Since the source and target sub-blocks lie in different logical rows, it then calls \(\text{SwapRows}\) to move the extracted bitmap from the lower row to the upper row. 
Its column offset from the target sub-block is 1, so the server calls $\text{Rot}_1$ to horizontally shift $vm_{kw_3}$ by 1 unit, aligning it to the target logical sub-block. 
Finally, in the homomorphic addition step, the server adds the aligned ciphertext into the target ciphertext. 
After completing the slot routing based on the public execution plan, bitmaps originally in different ciphertexts are routed to the same ciphertext, thereby providing a foundation for the subsequent efficient parallel execution of boolean operations. 

\begin{figure}[t]
	\centering
	\includegraphics[width=1.0\columnwidth]{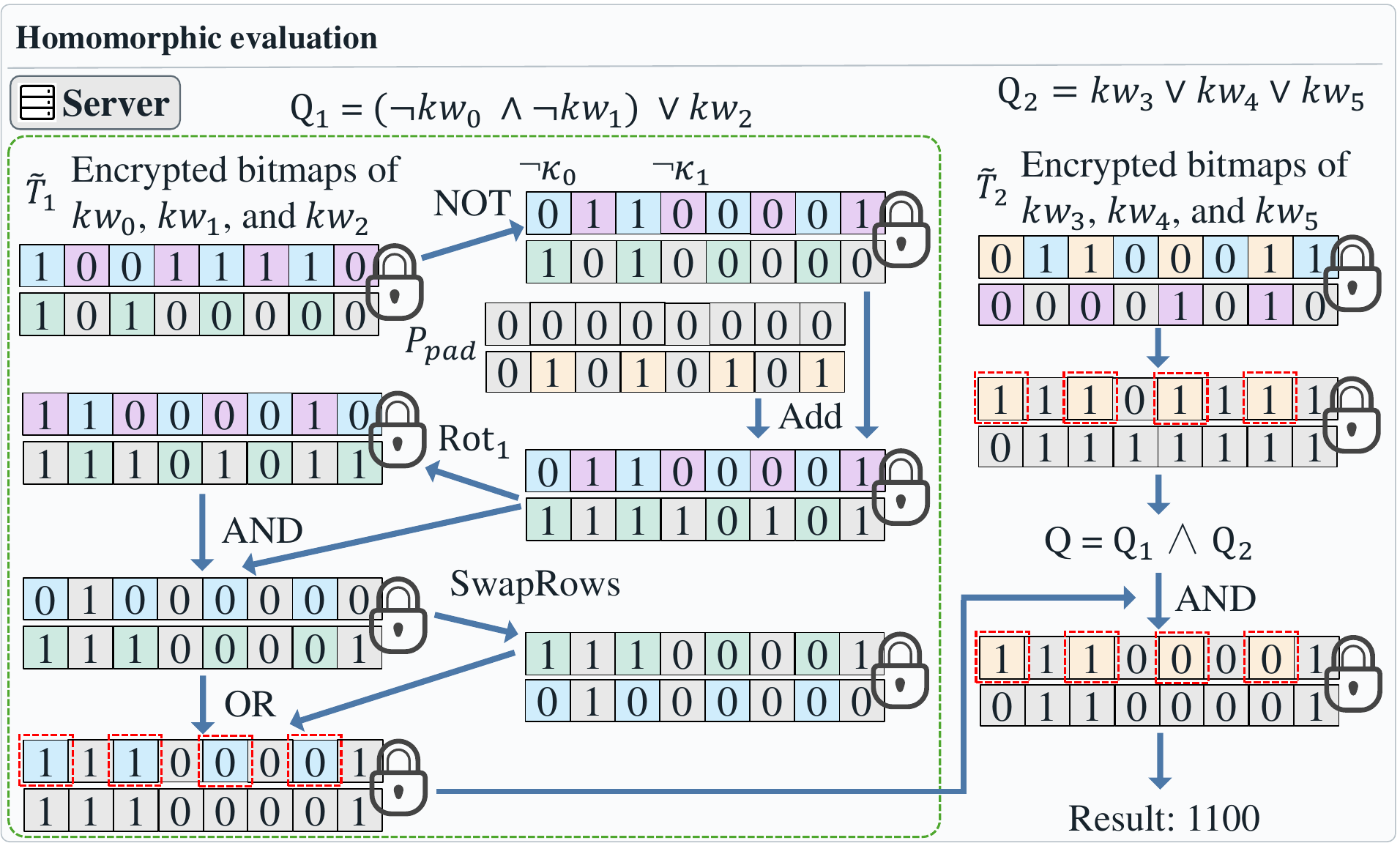}
	\vspace{-0.5cm}
	\caption{An example workflow for the homomorphic evaluation of boolean queries.}
	\label{fig:boolean}
	\vspace{-0.5cm}
\end{figure}

\subsection{Homomorphic Evaluation of Boolean Queries}
\label{subsec:Homomorphic Evaluation of Boolean Queries}

After aligning the ciphertexts via slot routing, the server performs the homomorphic evaluation of the boolean query bottom-up according to the public execution plan. 
For subtrees compiled into local DNFs, the server sequentially executes leaf node NOT processing, constant padding, intra-clause AND evaluation, and inter-clause OR evaluation. 
For unexpanded internal nodes, after computing their child nodes, the server evaluates them according to the original syntax tree. 

For subtrees compiled into local DNFs, the server processes the NOT operations pushed down to the leaf nodes. 
Since the slot values in the encrypted bitmap are all 0 or 1, the NOT operation can be directly represented as $\neg x=1-x$. 
Therefore, for leaf nodes with negation flags in the execution plan, the server subtracts the corresponding ciphertext from the constant 1. 
For leaf nodes without negation flags, it retains the original ciphertext. 
As shown in Fig. \ref{fig:boolean}, the server executes NOT operations on the corresponding leaf nodes of $Q_1$ in $\widetilde{T}_1$, obtaining $\neg kw_0$ and $\neg kw_1$. 
After completing the leaf node NOT operations, the server performs constant padding according to the public execution plan. 
The server constructs the padding plaintext $P_{pad}$ according to the following rules. 
Let a local DNF contain $C'$ clauses, and each clause occupy $L$ logical sub-blocks. 
Let the $i$-th actual clause contain $l_i$ conditions. 
When $l_i<L$, the server pads the remaining $L-l_i$ logical sub-blocks of this clause with all-one bitmaps. 
Since $x\land 1=x$, these padded values do not change the AND result of the actual clause. 
When the actual clause count $C<C'$, the client adds $C'-C$ virtual clauses into the public execution plan. 
The server keeps all $L$ logical sub-blocks of each virtual clause as all-zero bitmaps. 
Therefore, the AND result of the virtual clause is always 0, which satisfies $x \lor 0 = x$, thus it does not affect the subsequent OR operations. 
The server constructs the plaintext $P_{{pad}}$ according to the above rules, and computes $\widetilde{C} \leftarrow \operatorname{CtPtAdd}(\widetilde{C},P_{{pad}})$ to complete the constant padding. 
As shown in Fig. \ref{fig:boolean}, the target ciphertext corresponding to $Q_1$ is padded with all-one bitmaps according to the computation rules. 

After padding is complete, the server first computes the internal AND of each clause. 
Assuming each clause occupies $L$ consecutive logical sub-blocks, the $i$-th clause is located at $q_{iL}, q_{iL+1}, \ldots, q_{iL+L-1}$. 
The server aligns the bitmaps within the same clause in rounds of $\delta=1,2,4,\ldots,L/2$ and executes $\widetilde{C} \leftarrow \text{CtCtMul}(\widetilde{C}, \text{Rot}_{\delta}(\widetilde{C}))$. 
If the logical sub-blocks to be aligned are in different rows, it uses $\text{SwapRows}$ for alignment according to the public execution plan. 
After each ciphertext multiplication, the server calls $\text{Relinearize}$ to compress the resulting ciphertext to its initial length. 
After $\log_2L$ rounds of computation, the AND result of the $i$-th clause is stored in $q_{iL}$. 
Thus, the results of all clauses are located at $q_0, q_L, q_{2L}, \ldots, q_{(C'-1)L}$, respectively. 
The server computes the OR operation results of these outcomes using a pairwise merging approach. 
At the beginning of the $\rho$-th round, the intermediate results are located at $q_0, q_{2^\rho L}, q_{2 \cdot 2^\rho L}, \ldots$. 
For the $j$-th pair of results, the target logical sub-block is $q_d = q_{2j \cdot 2^\rho L}$ and the source logical sub-block is $q_s = q_{(2j+1) \cdot 2^\rho L}$, where $j=0,1,\ldots,C'/2^{\rho+1}-1$. 
The server first computes $X = \text{CtPtMul}(\widetilde{C}, E_{q_d})$ and $Z = \text{CtPtMul}(\widetilde{C}, E_{q_s})$ to extract the target result and source result, respectively. 
Subsequently, the server calculates the column offset $\Delta_c = (c_{q_s} - c_{q_d}) \bmod (N/2)$. 
If $r_{q_s} = r_{q_d}$, the server executes $W = \text{Rot}_{\Delta_c}(Z)$. 
If $r_{q_s} \neq r_{q_d}$, the server executes $W = \text{Rot}_{\Delta_c}(\text{SwapRows}(Z))$. 
The server then computes $X+W-\text{CtCtMul}(X,W)$ to obtain the OR operation result of the two clauses. 
The final result of the local DNF is stored in $q_0$. 
As shown in Fig.  \ref{fig:boolean}, the result of $Q_1$ is stored in the 0-th logical sub-block.

For unexpanded internal nodes, the client retains their original AND or OR operations in the public execution plan. 
The server processes these nodes bottom-up according to the query syntax tree. 
It only computes the result of an internal node after the evaluation of both its child nodes is completed. 
Let the encrypted bitmaps returned by the two child nodes be $X^\prime$ and $W^\prime$, respectively. 
When the internal node is AND, the server computes $Z^\prime = \text{CtCtMul}(X^\prime,W^\prime)$. 
When the internal node is OR, the server computes $Z^\prime = X^\prime+W^\prime-\text{CtCtMul}(X^\prime,W^\prime)$. 
The server repeats the above process until the evaluation of the root node of the query syntax tree is completed. 
As shown in Fig. \ref{fig:boolean}, after evaluating $Q_1$ and $Q_2$, the AND operation of the internal node is executed, yielding the encrypted result '1100', indicating that the first two documents satisfy the client's query. 

\subsection{PMKPIR}
\label{subsec:PMKPIR}

As illustrated in Fig. \ref{fig:system}, PMKPIR comprises two phases: offline initialization and online query. 
In the offline phase, the server executes \textbf{Setup} in Section \ref{subsec:details of PMKPIR} to generate the constant-weight codeword and inverted bitmap for each keyword in the database, and encodes them into BFV plaintexts according to the interleaved batching method in Section \ref{subsec:Interleaved Batching Method}. 

In the online query phase, the client first compiles the Boolean query according to Section \ref{subsec:query compilation}, extracts the queried keywords, and generates the public execution plan. 
Subsequently, the client executes \textbf{QueryGen} in Section \ref{subsec:details of PMKPIR} to encode every $B$ queried keywords into different logical sub-blocks of a BFV query ciphertext, and sends the query ciphertext alongside the public execution plan to the server. 
The server executes \textbf{Response} in Section \ref{subsec:details of PMKPIR} to compute the encrypted inverted bitmaps corresponding to each queried keyword in parallel during a single database traversal. 
Subsequently, based on the public query plan sent by the client, the server completes slot routing according to Section \ref{subsec:slot routing}. 
It then performs homomorphic evaluation according to Section \ref{subsec:Homomorphic Evaluation of Boolean Queries}, sequentially computing each local DNF and the internal nodes retained in the syntax tree. 
Ultimately, this process yields the encrypted document selection bitmap corresponding to the query conditions. 
The server returns the result ciphertexts to the client. 
The client decrypts each result ciphertext and concatenates them in the order of the bitmap chunks to recover the complete document selection bitmap, where the positions with a value of 1 correspond to the document IDs that satisfy the boolean query conditions. 

After obtaining the matched document indices from the decrypted bitmap, the subsequent retrieval of document content is handled by an independent payload recovery phase. 
Let the document size be $S_\ell$, and the backend encoding capacity per document be $P_{cap}$. 
When $S_\ell \le P_{cap}$, the system adopts Multi-query PIRANA \cite{LiuLWR24} for batch retrieval. 
When $S_\ell > P_{cap}$, it adopts SpiralStreamPack \cite{MenonW22} for chunked streaming transmission. 
The payload recovery backend is determined during initialization based on $S_\ell$, which does not alter the core PMKPIR protocol flow. 

\section{Security Sketch}

We consider an honest-but-curious server that executes the protocol correctly but attempts to infer the queried keywords from the query ciphertexts, intermediate ciphertexts, and response ciphertexts observed during protocol execution. 
For IBPIR, let $KL$ denote a queried keyword list and let $\alpha_{KL}=\lceil |KL|/B\rceil$ denote the number of query ciphertexts. 
For any two keyword lists $KL_0$ and $KL_1$ satisfying $\alpha_{KL_0}=\alpha_{KL_1}$, if the server cannot distinguish the views generated by $KL_0$ and $KL_1$, IBPIR is said to satisfy indistinguishability under chosen keyword attack (IND-CKA) \cite{SongXZZLMX26}. 
PMKPIR adopts the same query-privacy boundary as BKPIR \cite{SongXZZLMX26}. 
In BKPIR, indistinguishability under chosen Boolean query attack (IND-CBQA) is defined for boolean queries with the same query arity and logical structure, thereby treating this structural information as public. 
In PMKPIR, this public information is represented explicitly by the execution plan. 
Specifically, let \(\Pi_Q\) denote the public execution plan of a boolean query \(Q\), let \(\alpha_Q\) denote the number of query ciphertexts, and define the leakage function as $\mathcal{L}(Q)=(\Pi_Q,\alpha_Q)$. 
The execution plan reveals the compiled boolean structure and query arity, but does not contain the queried keyword values. 
For any two boolean queries satisfying \(\mathcal{L}(Q_0)=\mathcal{L}(Q_1)\), PMKPIR satisfies IND-CBQA if the server cannot distinguish their corresponding protocol views. 
Assuming the hardness of the RLWE problem and the IND-CPA security of BFV, IBPIR is IND-CKA secure and PMKPIR is IND-CBQA secure under $\mathcal{L}$. 
In both schemes, the queried keywords are transmitted exclusively as BFV ciphertexts. 
IBPIR performs keyword matching through public homomorphic operations, while PMKPIR further performs slot routing and boolean evaluation according to the public execution plan. 
Queries with the same public leakage induce the same execution flow, while the specific queried keywords remain hidden by BFV encryption. 
Appendix \ref{app:security} provides the formal security proofs, analyzes response privacy, execution plan leakage, and security under database updates, and discusses optional padding mechanisms for mitigating query structure leakage. 

\section{Evaluation}

We conduct our experiments on a single desktop workstation equipped with an Intel(R) Core(TM) i9-14900K CPU (3.20 GHz, 24 cores, 32 threads), 96 GB of DDR5 memory (95.8 GB available), and a 930 GB KIOXIA-EXCERIA PRO SSD. 
All experiments are executed in WSL2, which exposes 46 GiB of memory and 12 GiB of swap space to the Linux environment. 

\subsection{Implementation}

We implement the PMKPIR system prototype using C++20 and conduct the experimental evaluation in a WSL2 environment. 
The homomorphic encryption is based on the BFV scheme from the Microsoft SEAL 4.1.2 library \cite{seal40}. 
The prototype adopts a polynomial modulus degree $N=16384$ and a plaintext modulus $t=65537$, with the coefficient modulus selected according to the 'tc128' standard of Microsoft SEAL to satisfy the 128-bit security level. 
The system's payload recovery backend is configured during the initialization phase based on the public document size. 
Since this prototype processes fixed-length documents of 64 bytes and the expected query results exhibit sparsity, we adopt the low-storage streaming Multi-query PIRANA \cite{LiuLWR24} with a fixed capacity of $K_0=32$ as the backend implementation. 

\subsection{Experimental Setup}

\noindent \textbf{Baselines.}
We compare PMKPIR with two representative baseline schemes: 
1) BKPIR \cite{SongXZZLMX26}, the state-of-the-art PIR scheme supporting universal boolean retrieval, and 
2) MkPIR \cite{ZhaoLHMZLCZ26}, the state-of-the-art multi-keyword PIR scheme that only supports conjunctive queries, including its basic version $\text{MkPIR}^{I}$ and the amortized version $\text{MkPIR}^{I*}$. 
We implement both BKPIR and MkPIR based on their respective open-source code. 
To ensure a fair comparison, all schemes use the same datasets, queried keywords, BFV parameters, and security level. 
All implementations are executed with a single thread bound to the same CPU core, ensuring that the measured performance differences arise from the protocols rather than multi-core parallelism or thread scheduling. 
Except for the end-to-end evaluation, all experiments use the client's recovery of the correct set of matched document IDs as the common endpoint. 
Since MkPIR supports only multi-keyword conjunctive queries, its two variants participate only in experiments with purely conjunctive workloads. 

\noindent \textbf{Default configurations.}
Unless otherwise specified, all experiments adopt the following default configurations. 
The database contains $m=2^{10}$ documents and $n=2^6$ distinct keywords, with each document containing $doc=2^2$ keywords. 
The query consists of $s=2^2$ distinct keywords, adopting a four-keyword conjunctive structure by default, and the number of matched documents is set to $r=2^3$. 
Each query ciphertext of PMKPIR is divided into $B=2^2$ logical sub-blocks, the length of the constant-weight coding is $w=2^{12}$, the Hamming weight is set to $k=2$, and the minimum remaining noise budget requirement is 20 bits. 

\noindent \textbf{Evaluation metrics.} 
We primarily adopt query time, communication overhead, peak memory, and storage overhead as evaluation metrics. 
The query time includes the time required from the client generating the query ciphertext to the client obtaining the retrieved set of document IDs. 
The communication overhead includes the query ciphertext and public execution plan sent by the client, as well as the result ciphertext returned by the server. 
The peak memory is the maximum memory usage during the query process. 
The storage overhead refers to the size of the encoded database on the server.

\subsection{System Performance Evaluation}

\begin{figure}[t]
	\centering
	\includegraphics[width=\linewidth]{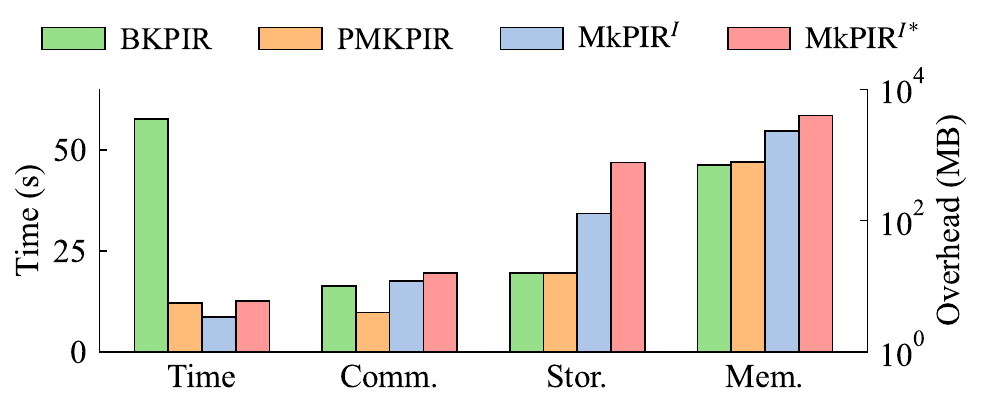}
	\vspace{-0.6cm}
	\caption{(Exp\#1) Baseline performance evaluation.}
	\label{fig:exp1}
	\vspace{-0.5cm}
\end{figure}

\noindent \textbf{Exp\#1 (Baseline performance evaluation).} 
We evaluate the query time, communication overhead, storage overhead, and peak memory under the default configuration. 
As shown in Fig. \ref{fig:exp1}, the query time of PMKPIR is reduced by 79.06\% compared to BKPIR, and the communication overhead is reduced by 60.00\%. 
The storage overhead of both schemes is identical, and their peak memory is comparable. 
Compared with PMKPIR, $\text{MkPIR}^{I}$ exhibits lower query time, but its communication overhead, storage overhead, and peak memory increase by $2.00\times$, $7.00\times$, and $1.96\times$, respectively. 
The query time of $\text{MkPIR}^{I*}$ increases by 4.68\%, and its communication overhead, storage overhead, and peak memory increase by $3.00\times$, $46.98\times$, and $4.07\times$, respectively. 
In addition, the minimum remaining noise budget of all schemes is no less than 210 bits, ensuring correct BFV decryption. 
Overall, PMKPIR significantly reduces the query time and communication overhead of BKPIR, and outperforms the two MkPIR schemes in terms of communication, storage, and peak memory. 
This is because PMKPIR encodes multiple queried keywords into the same ciphertext and performs multiple keyword matchings in parallel during a single database traversal, thereby reducing computational and communication overheads. 

\begin{figure}[t]
	\centering
	
	
	\includegraphics[width=0.8\linewidth]{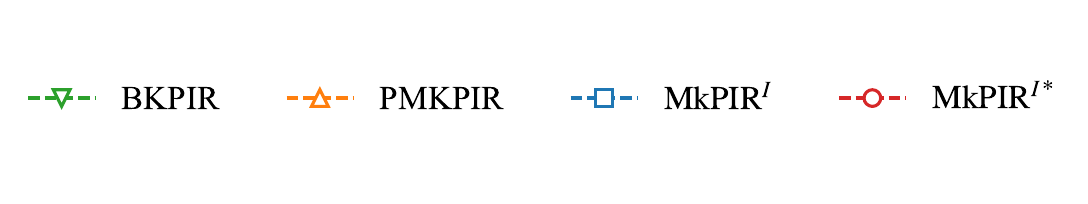} \\
	\vspace{-0.5cm} 
	
	\begin{subfigure}[b]{0.22\textwidth}
		\centering
		\includegraphics[width=\textwidth]{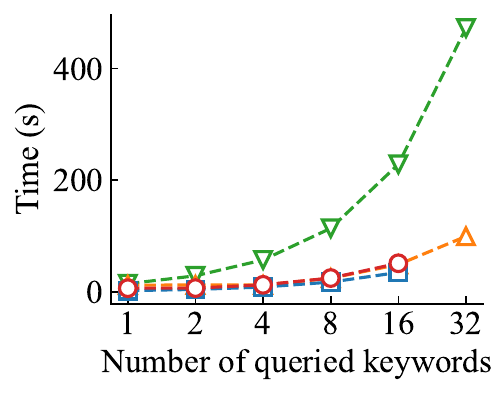}
		\vspace{-0.6cm}
		\caption{Query time}
		\label{fig:exp2.1}
	\end{subfigure}
	\hfill 
	\begin{subfigure}[b]{0.22\textwidth}
		\centering
		\includegraphics[width=\textwidth]{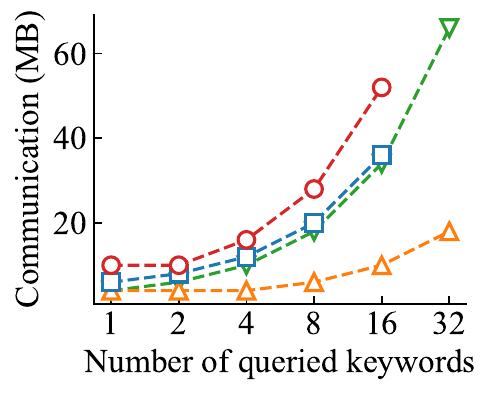}
		\vspace{-0.6cm}
		\caption{Communication overhead}
		\label{fig:exp2.2}
	\end{subfigure}
	
	
	\begin{subfigure}[b]{0.22\textwidth}
		\centering
		\includegraphics[width=\textwidth]{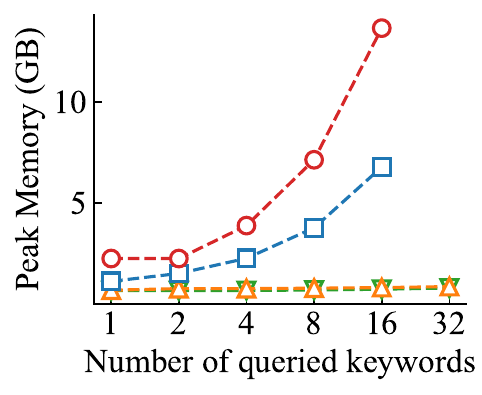}
		\vspace{-0.6cm}
		\caption{Peak memory}
		\label{fig:exp2.3}
	\end{subfigure}
	\hfill
	\begin{subfigure}[b]{0.22\textwidth}
		\centering
		\includegraphics[width=\textwidth]{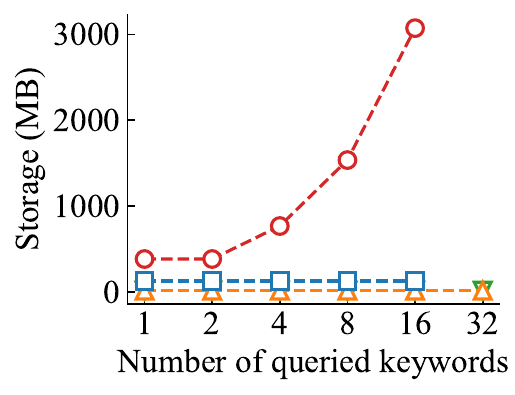}
		\vspace{-0.6cm}
		\caption{Storage overhead}
		\label{fig:exp2.4}
	\end{subfigure}
	
	\vspace{-0.1cm}
	
	\caption{(Exp\#2) Impact of the number of queried keywords.}
	\label{fig:exp2}
	\vspace{-0.6cm}
\end{figure}

\noindent \textbf{Exp\#2 (Impact of the number of queried keywords).} 
As shown in Fig. \ref{fig:exp2}, when the number of queried keywords increases from 1 to 32, PMKPIR reduces the query time by 17.97\%–78.96\% and the communication overhead by up to 72.73\% compared with BKPIR. 
At 32 queried keywords, both MkPIR variants fail because the Gödel product used for set intersection cannot be represented under the fixed plaintext modulus \(t\)=65537, whereas PMKPIR and BKPIR successfully complete the query. 
These results demonstrate that interleaved batching improves query and communication efficiency as the number of queried keywords increases. 

\begin{figure}[t]
	\centering
	
	\vspace{-0.4cm}
	
	\includegraphics[width=0.8\linewidth]{images/legend_only.pdf} \\
	\vspace{-0.5cm} 
	
	\begin{subfigure}[b]{0.22\textwidth}
		\centering
		\includegraphics[width=\textwidth]{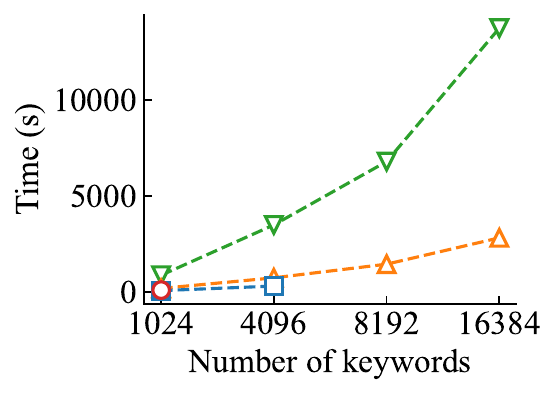}
		\vspace{-0.6cm}
		\caption{Query time}
		\label{fig:exp3.1}
	\end{subfigure}
	\hfill 
	\begin{subfigure}[b]{0.22\textwidth}
		\centering
		\includegraphics[width=\textwidth]{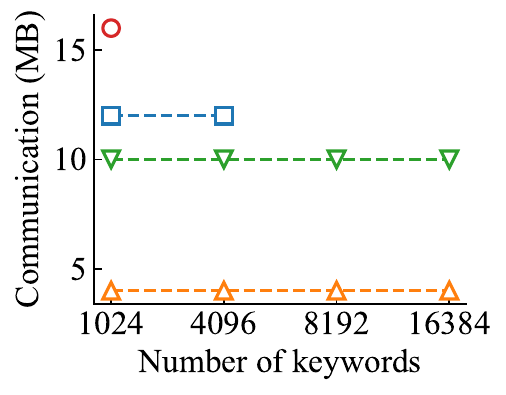}
		\vspace{-0.6cm}
		\caption{Communication overhead}
		\label{fig:exp3.2}
	\end{subfigure}
	
	
	\begin{subfigure}[b]{0.22\textwidth}
		\centering
		\includegraphics[width=\textwidth]{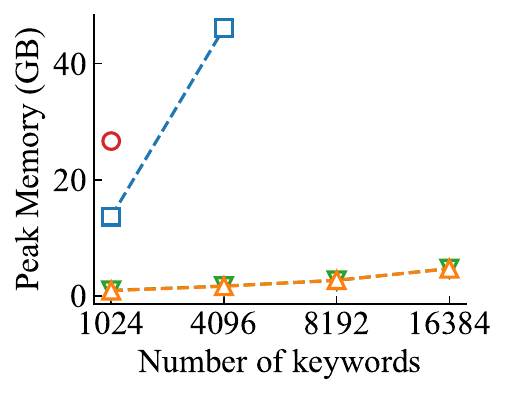}
		\vspace{-0.6cm}
		\caption{Peak memory}
		\label{fig:exp3.3}
	\end{subfigure}
	\hfill
	\begin{subfigure}[b]{0.22\textwidth}
		\centering
		\includegraphics[width=\textwidth]{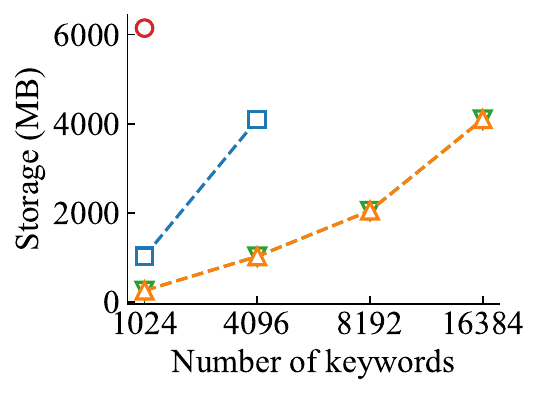}
		\vspace{-0.6cm}
		\caption{Storage overhead}
		\label{fig:exp3.4}
	\end{subfigure}
	
	\vspace{-0.1cm}
	
	\caption{(Exp\#3) Impact of the number of keywords.}
	\label{fig:exp3}
	\vspace{-0.6cm}
\end{figure}

\begin{table*}[t]
	\centering
	\caption{(Exp\# 4) Performance under different Boolean query structures.}
	\vspace{-0.1cm}
	\label{tab:exp4}
	\small
	\begin{tabular*}{\textwidth}{@{\extracolsep{\fill}} llccccc @{}}
		\toprule
		Query structure
		& Scheme
		& \makecell{Query\\time (s)}
		& \makecell{Communication\\(MB)}
		& \makecell{Peak memory\\(GB)}
		& \makecell{Storage\\(MB)}
		& \makecell{Minimum noise\\(bits)} \\
		\midrule
		
		AND
		& BKPIR \cite{SongXZZLMX26}
		& 108.47 & 18.00 & 0.74 & 16.01 & 181 \\
		& PMKPIR
		& 23.21 & 6.00 & 0.71 & 16.01 & 181 \\
		
		\addlinespace
		OR
		& BKPIR \cite{SongXZZLMX26}
		& 107.07 & 18.00 & 0.74 & 16.01 & 181 \\
		& PMKPIR
		& 23.11 & 6.00 & 0.72 & 16.01 & 181 \\
		
		\addlinespace
		Balanced
		& BKPIR \cite{SongXZZLMX26}
		& 108.30 & 18.00 & 0.74 & 16.01 & 181 \\
		& PMKPIR
		& 23.05 & 6.00 & 0.72 & 16.01 & 182 \\
		
		\addlinespace
		Left-deep
		& BKPIR \cite{SongXZZLMX26}
		& 108.03 & 18.00 & 0.73 & 16.01 & 65 \\
		& PMKPIR
		& 23.08 & 6.00 & 0.71 & 16.01 & 65 \\
		
		\addlinespace
		NOT
		& BKPIR \cite{SongXZZLMX26}
		& 108.36 & 18.00 & 0.74 & 16.01 & 181 \\
		& PMKPIR
		& 23.33 & 6.00 & 0.77 & 16.01 & 183 \\
		
		\bottomrule
	\end{tabular*}
	\vspace{-0.3cm}
\end{table*}

\noindent \textbf{Exp\#3 (Impact of the number of keywords).} 
We increase the number of keywords $n$ in the database from 1,024 to 16,384. 
As shown in Fig. \ref{fig:exp3}, compared to BKPIR, the query time of PMKPIR is reduced by 78.69\%–79.48\%, and the communication overhead is reduced by 60.00\%, while the storage overhead and peak memory of both schemes are comparable. 
Although $\text{MkPIR}^{I}$ exhibits lower query time under successfully executed configurations, its communication overhead, storage overhead, and peak memory increase by $2.00\times$, $3.00\times$, and $13.64\times$–$26.38\times$, respectively; when $n=4,096$, its peak memory reaches 46.08 GB, and it fails to complete the experiment starting from $n=8,192$. 
The peak memory of $\text{MkPIR}^{I*}$ already reaches 26.65 GB when $n=1,024$. 
As $n$ increases, the intermediate results generated by homomorphic computations expand rapidly; therefore, it fails to complete the experiment starting from $n=4,096$. 
The minimum remaining noise budget of all successfully executed configurations is no less than 207 bits. 
The experimental results demonstrate that as the number of distinct keywords increases, PMKPIR exhibits stable computation, communication, and memory efficiency.

\noindent \textbf{Exp\#4 (Impact of boolean query structure).} 
To evaluate the impact of different query structures on system performance, we fix the query statement to contain 8 distinct keywords and construct 5 typical boolean query scenarios: all-AND, all-OR, balanced tree, left-deep tree, and queries containing NOT operations. 
We then evaluate their impact on query time, communication overhead, peak memory, and storage overhead. 
As shown in Table \ref{tab:exp4}, under different boolean query structures, the query time of PMKPIR is reduced by 78.42\%–78.72\% compared to BKPIR, and the communication overhead is reduced by 66.67\%. 
The storage overhead of both schemes is identical, and their peak memory is comparable. 
The fluctuation in the query time of PMKPIR across the five query structures does not exceed 1.23\%, indicating that the query structure has a minor impact on the overall query time. 
The minimum remaining noise budget across all configurations is no less than 65 bits, ensuring correct BFV decryption. 
Experimental results demonstrate that PMKPIR can correctly and efficiently support AND, OR, and NOT operations, as well as boolean syntax trees of different shapes. 
It should be noted that since the two MkPIR schemes only support conjunctive queries, they are excluded from this experiment. 

\begin{figure}[t]
	\centering
	
	\vspace{-0.4cm}
	
	\includegraphics[width=0.8\linewidth]{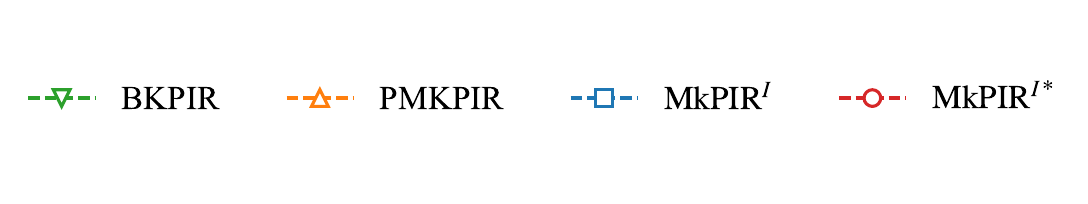} \\
	\vspace{-0.5cm} 
	
	\begin{subfigure}[b]{0.22\textwidth}
		\centering
		\includegraphics[width=\textwidth]{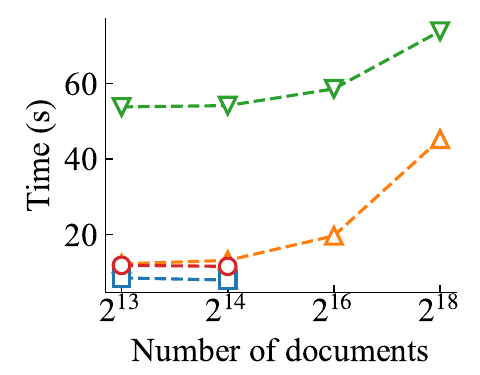}
		\vspace{-0.6cm}
		\caption{Query time}
		\label{fig:exp5.1}
	\end{subfigure}
	\hfill 
	\begin{subfigure}[b]{0.22\textwidth}
		\centering
		\includegraphics[width=\textwidth]{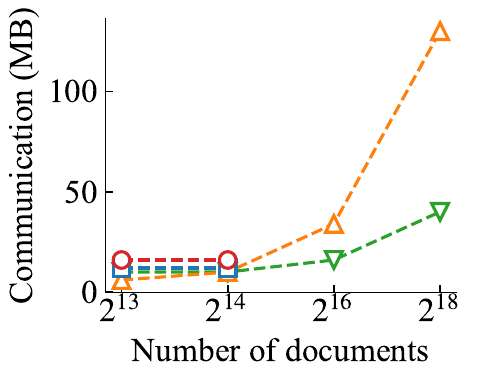}
		\vspace{-0.6cm}
		\caption{Communication overhead}
		\label{fig:exp5.2}
	\end{subfigure}
	
	
	\begin{subfigure}[b]{0.22\textwidth}
		\centering
		\includegraphics[width=\textwidth]{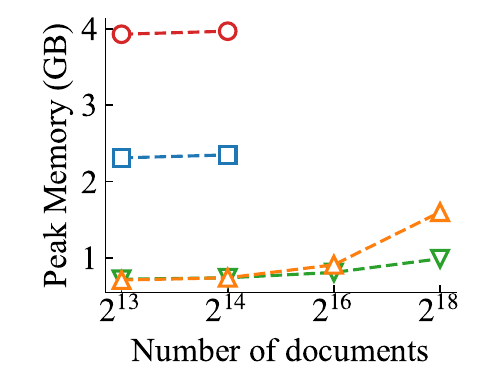}
		\vspace{-0.6cm}
		\caption{Peak memory}
		\label{fig:exp5.3}
	\end{subfigure}
	\hfill
	\begin{subfigure}[b]{0.22\textwidth}
		\centering
		\includegraphics[width=\textwidth]{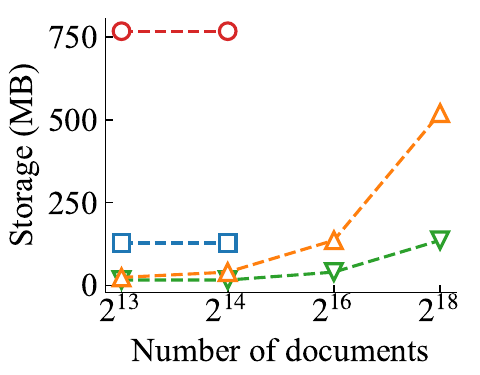}
		\vspace{-0.6cm}
		\caption{Storage overhead}
		\label{fig:exp5.4}
	\end{subfigure}
	
	\vspace{-0.1cm}
	
	\caption{(Exp\#5) Impact of the number of documents.}
	\label{fig:exp5}
	\vspace{-0.6cm}
\end{figure}

\noindent \textbf{Exp\#5 (Impact of the number of documents).} 
We evaluate the impact of increasing the number of documents $m$ from \(2^{13}\) to \(2^{18}\). 
As shown in Fig. \ref{fig:exp5}, PMKPIR reduces the query time by 38.71\%–77.22\% compared with BKPIR. 
As bitmap chunking produces more response ciphertexts, its communication and storage overheads increase with \(m\), while the peak memory remains below 1.60 GB. 
Although the two MkPIR schemes are faster at \(m\le 2^{14}\), they incur substantially higher storage and peak memory overheads and fail from \(m=2^{16}\) because their encoded results exceed the available BFV slots. 
All successful configurations retain at least 210 bits of noise budget. 
Overall, PMKPIR provides lower query time than BKPIR while supporting substantially more documents than the MkPIR schemes. 

\begin{table}[t]
	\centering
	\caption{(Exp\#6) Offline phase overhead.}
	\vspace{-0.1cm}
	\label{tab:exp6}
	\small
	\begin{tabularx}{\linewidth}{@{} l >{\centering\arraybackslash}X >{\centering\arraybackslash}X >{\centering\arraybackslash}X @{}}
		\toprule
		Scheme 
		& \makecell{KeyGen \\ (s)} 
		& \makecell{DB Prep \\ (s)} 
		& \makecell{Eval Keys \\ (MB)} \\
		\midrule
		BKPIR \cite{SongXZZLMX26}                & 0.64 & 0.01 & 270.14 \\
		PMKPIR                                   & 0.67 & 0.02 & 288.14 \\
		$\text{MkPIR}^{I}$ \cite{ZhaoLHMZLCZ26}  & 0.65 & 0.20 & 270.14 \\
		$\text{MkPIR}^{I*}$ \cite{ZhaoLHMZLCZ26} & 0.98 & 1.00 & 270.14 \\
		\bottomrule
	\end{tabularx}
\end{table}

\begin{table*}[t]
	\centering
	\caption{(Exp\#10) Theoretical complexity comparison.}
	\label{tab:exp10}
	\small
	\begin{threeparttable} 
		\begin{tabular*}{\textwidth}{@{\extracolsep{\fill}} lcccc @{}}
			\toprule
			Server operation 
			& BKPIR 
			& PMKPIR 
			& $\text{MkPIR}^{I}$ 
			& $\text{MkPIR}^{I*}$ \\
			\midrule
			
			Keyword matching
			& $O(sn(k+\log N+f_N))$
			& $O(\alpha n(k+\log w+f_B))$
			& $O(s\log n)$
			& $O(\chi\log\mu)$ \\
			
			\addlinespace
			Slot routing
			& --
			& $O(f_BLC')$
			& --
			& -- \\
			
			\addlinespace
			Boolean evaluation
			& $O(f_NG_Q)$
			& $O(f_B(\log L+C'+G_{\mathrm{AST}}))$
			& $O(s)$
			& $O(\chi)$ \\
			
			\addlinespace
			Relinearization and key switching
			& \makecell{$O\big(sn(k+\log N)$ \\ $+f_NG_Q\big)$}
			& \makecell{$O\big(\alpha n(k+\log w)$ \\ $+f_B(LC'+\log L+C'+G_{\mathrm{AST}})\big)$}
			& $O(s\log n)$
			& $O(\chi\log\mu)$ \\
			
			\addlinespace
			Multiplicative depth
			& $\lceil\log_2k\rceil+D_Q$
			& \makecell{$\lceil\log_2k\rceil+\log_2L$ \\ $+\log_2C'+D_{\mathrm{AST}}$}
			& $\lceil\log_2s\rceil$
			& $\lceil\log_2\chi\rceil$ \\
			
			\addlinespace
			Response ciphertexts
			& $f_N$
			& $f_B$
			& $s+1$
			& $\chi+1$ \\
			\bottomrule
		\end{tabular*}
		
		\begin{tablenotes}[flushleft] 
			\footnotesize
			\item \emph{Note.} 
			$f_N=\lceil m/N\rceil$ and $f_B=\lceil m/w\rceil$ denote the numbers of bitmap chunks in BKPIR and PMKPIR, respectively. 
			$G_Q$ and $D_Q$ denote the number and multiplicative depth of boolean operations in BKPIR; $G_{\mathrm{AST}}$ and $D_{\mathrm{AST}}$ denote the number and multiplicative depth of the retained AST operations in PMKPIR. 
			For $\text{MkPIR}^{I*}$, $\chi$ and $\mu$ denote the number of bucket queries and the partition size, respectively.
		\end{tablenotes}
	\end{threeparttable} 
\end{table*}

\begin{figure}[t]
	\centering
	
	\vspace{-0.4cm}
	
	\includegraphics[width=0.8\linewidth]{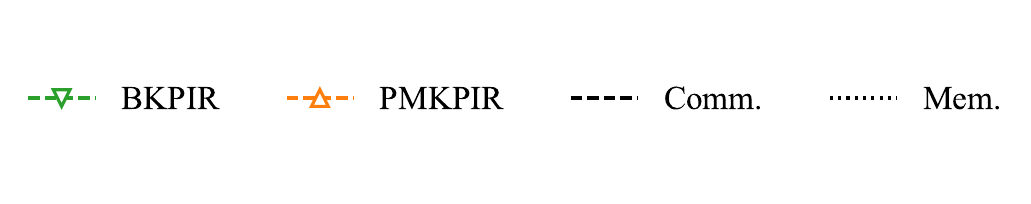} \\
	\vspace{-0.6cm} 
	
	\begin{subfigure}[t]{0.22\textwidth}
		\centering
		\includegraphics[width=\textwidth]{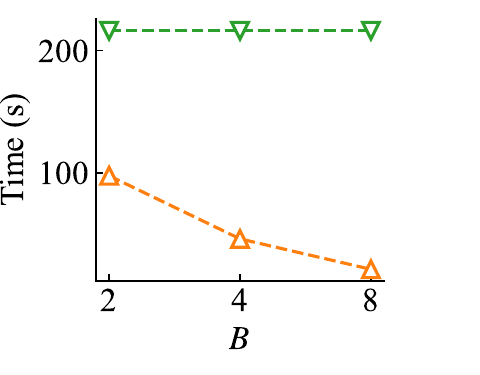}
		\vspace{-0.6cm}
		\caption{Query time}
		\label{fig:exp7.1}
	\end{subfigure}
	\hfill 
	\begin{subfigure}[t]{0.22\textwidth}
		\centering
		\includegraphics[width=\textwidth]{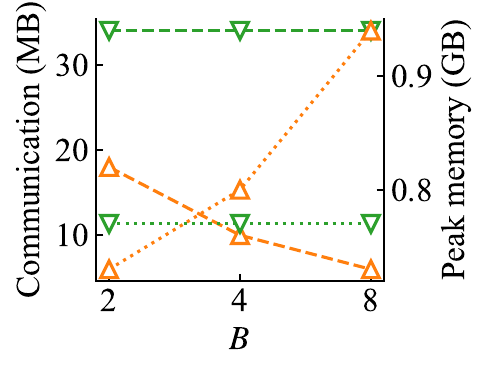}
		\vspace{-0.6cm}
		\caption{Communication overhead and peak memory}
		\label{fig:exp7.2}
	\end{subfigure}
	
	\vspace{-0.1cm}
	
	\caption{(Exp\#7) Impact of the number of logical sub-blocks $B$.}
	\label{fig:exp7}
	\vspace{-0.6cm}
\end{figure}

\noindent \textbf{Exp\#6 (Offline phase overhead).} 
We measure database preprocessing, including keyword encoding and index construction, and BFV key generation under the default configuration. 
We also measure the storage of evaluation keys, including the relinearization and Galois keys used for relinearization and slot rotation. 
As shown in Table \ref{tab:exp6}, the offline overhead of PMKPIR primarily stems from key generation. 
Its total time increases by 5.08\% compared to BKPIR, but is reduced by 19.42\%--65.20\% compared to the two MkPIR schemes. 
The evaluation keys of PMKPIR occupy 288.14 MB, which is 6.66\% higher than the 270.14 MB required by the other schemes. 
This increment is attributed to the additional rotation keys required for interleaved batching and slot routing.

\noindent \textbf{Exp\#7 (Impact of the number of logical sub-blocks $B$).} 
We vary \(B\) from 2 to 8 and use BKPIR as the non-batched baseline. 
As shown in Fig. \ref{fig:exp7}, increasing \(B\) from 2 to 8 further reduces the query time and communication overhead by 77.76\% and 66.67\%, respectively, while increasing peak memory by 28.81\%. 
A larger \(B\) packs more queried keywords into each query ciphertext, thereby reducing the number of query ciphertexts and database traversals at the cost of higher peak memory.

\begin{figure}[t]
	\centering
	
	\vspace{-0.7cm}
	
	\includegraphics[width=0.85\linewidth]{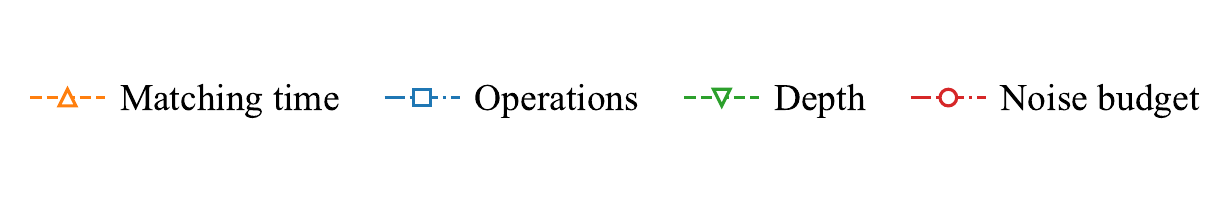} \\
	\vspace{-0.5cm} 
	
	\begin{subfigure}[b]{0.22\textwidth}
		\centering
		\includegraphics[width=\textwidth]{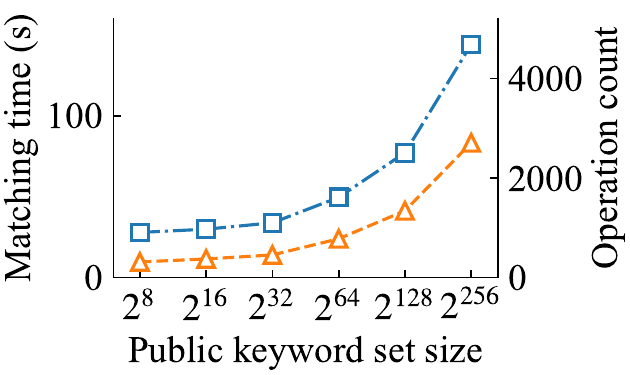}
		\vspace{-0.5cm}
		\caption{Keyword matching time and the number of homomorphic operations}
		\label{fig:exp8.1}
	\end{subfigure}
	\hfill 
	\begin{subfigure}[b]{0.22\textwidth}
		\centering
		\includegraphics[width=\textwidth]{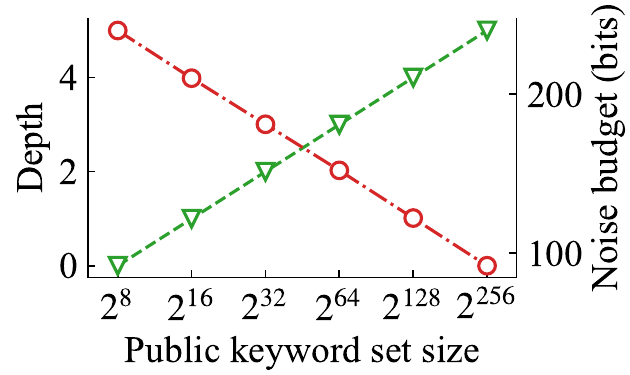}
		\vspace{-0.5cm}
		\caption{Multiplicative depth and minimum remaining noise budget}
		\label{fig:exp8.2}
	\end{subfigure}
	
	\vspace{-0.1cm}
	
	\caption{(Exp\#8) Impact of the size of the public keyword set.}
	\label{fig:exp8}
	\vspace{-0.7cm}
\end{figure}

\noindent \textbf{Exp\#8 (Impact of the size of the public keyword set).} 
We evaluate the impact of the public keyword set size \(\lvert KW\rvert\) on the keyword matching time, number of homomorphic operations, multiplicative depth, and minimum remaining noise budget of PMKPIR. 
As \(\lvert KW\rvert\) increases from \(2^8\) to \(2^{256}\), the minimum Hamming weight \(k\) satisfying \(\binom{w}{k}\geq\lvert KW\rvert\) increases from 1 to 31. 
As shown in Fig. \ref{fig:exp8}, the keyword matching time and number of homomorphic operations increase by 8.67\(\times\) and 5.17\(\times\), respectively, while the multiplicative depth increases from 0 to 5. 
This is because a larger \(k\) requires more homomorphic operations for equality evaluation, whereas \(\text{EvalTree}\) limits the multiplicative depth to logarithmic growth. 
The minimum remaining noise budget decreases from 240 to 92 bits, and all configurations are successfully decrypted. 
Since \(\lvert KW\rvert\) does not affect the number of query or response ciphertexts, the communication overhead remains unchanged. 

\noindent \textbf{Exp\#9 (End-to-end evaluation).} 
The end-to-end overhead consists of matched document ID recovery and document content retrieval. 
Compared to BKPIR and \(\text{MkPIR}^{I*}\), the end-to-end time of PMKPIR is reduced by 78.14\% and 1.98\%, respectively, whereas it increases by 40.53\% compared to \(\text{MkPIR}^{I}\). 
Its end-to-end communication overhead is reduced by 35.03\%, 41.83\%, and 51.89\% compared to BKPIR, \(\text{MkPIR}^{I}\) and \(\text{MkPIR}^{I*}\), respectively, while maintaining the lowest peak memory. 
Although \(\text{MkPIR}^{I}\) achieves the shortest end-to-end time, it incurs substantially higher communication and memory overheads and supports only conjunctive queries rather than universal boolean computation. 
Overall, PMKPIR achieves a better balance among end-to-end latency, communication overhead, and peak memory consumption. 

\noindent \textbf{Exp\#10 (Theoretical complexity).} 
Table \ref{tab:exp10} shows that interleaved batching reduces the keyword matching factor of PMKPIR from \(s\) to \(\alpha=\lceil s/B\rceil\). 
Its additional slot routing and boolean evaluation costs depend on the bitmap chunks and query structure rather than \(n\), and can therefore be amortized when keyword matching dominates. 
MkPIR has lower matching complexity, but only supports conjunctive queries and returns \(s+1\) or \(\chi+1\) ciphertexts, whereas PMKPIR supports universal boolean computation with \(f_B\) response ciphertexts. 

\begin{table}[t]
	\centering
	\vspace{-0.3cm}
	\caption{(Exp\#9) End-to-end evaluation.}
	\label{tab:exp9}
	\small
	\begin{tabularx}{\linewidth}{@{} l >{\centering\arraybackslash}X >{\centering\arraybackslash}X >{\centering\arraybackslash}X @{}}
		\toprule
		Scheme 
		& \makecell{End-to-end \\ (s)} 
		& \makecell{Comm. \\ (MB)} 
		& \makecell{Peak memory \\ (MB)} \\
		\midrule
		BKPIR \cite{SongXZZLMX26}                & 55.13 & 17.13 & 759.49  \\
		PMKPIR                                   & 12.05 & 11.13 & 720.45  \\
		$\text{MkPIR}^{I}$ \cite{ZhaoLHMZLCZ26}  & 8.58  & 19.13 & 2406.03 \\
		$\text{MkPIR}^{I*}$ \cite{ZhaoLHMZLCZ26} & 12.30 & 23.13 & 4067.02 \\
		\bottomrule
	\end{tabularx}
	\vspace{-0.4cm}
\end{table}

\section{Conclusion}

We propose IBPIR, a multi-keyword PIR scheme based on interleaved batching. 
It addresses the limitation of existing private boolean retrieval schemes that require repeated database traversals per queried keyword. 
By encoding multiple queried keywords into a single BFV ciphertext, IBPIR enables parallel multi-keyword matching during a single database traversal, thereby reducing both query time and communication overhead. 
Building upon this, we further propose PMKPIR, which accommodates universal boolean computation while preserving the performance advantages of interleaved batching, providing a novel approach for efficient private boolean retrieval.

\bibliographystyle{IEEEtran}
\bibliography{IEEEabrv,bib}

\appendix
\label{app:security}

\subsection{Security Proofs}

\noindent \textbf{Definition 1 (IND-CKA).} 
IBPIR satisfies IND-CKA if for any probabilistic polynomial-time (PPT) adversary $\mathcal{A}$, and any two keyword lists 
$$
\begin{aligned}
	KL_0 &= (kw_0^{(0)},\ldots,kw_{s_0-1}^{(0)}), \\
	KL_1 &= (kw_0^{(1)},\ldots,kw_{s_1-1}^{(1)})
\end{aligned}
$$
satisfying $\alpha_{KL_0}=\alpha_{KL_1}$, the following condition holds:
$$ 
\begin{aligned}
	\Big| \Pr[\mathcal{A}(\mathsf{View}_{\mathrm{I}}(KL_0))=1] &
	- \Pr[\mathcal{A}(\mathsf{View}_{\mathrm{I}}(KL_1))=1] \Big| &  \\\leq \mathsf{negl}(\lambda). 
\end{aligned}
$$
Here, $\lambda$ is the security parameter, and $\alpha_{KL}=\lceil |KL|/B \rceil$ denotes the number of query ciphertexts. Furthermore, $\mathsf{View}_{\mathrm{I}}(KL)$ contains all query ciphertexts, homomorphic computation traces, and response ciphertexts generated during the protocol execution. 

\noindent \textbf{Definition 2 (IND-CBQA).}
Let $\Pi_{\mathcal{Q}}$ denote the public execution plan of a boolean query $\mathcal{Q}$, and $\alpha_{\mathcal{Q}}$ denote the number of query ciphertexts. 
We define the leakage function as $\mathcal{L}(\mathcal{Q}) = (\Pi_{\mathcal{Q}}, \alpha_{\mathcal{Q}}).$
The PMKPIR protocol satisfies IND-CBQA if for any PPT adversary $\mathcal{A}$, and any two boolean queries $\mathcal{Q}_0$ and $\mathcal{Q}_1$ satisfying $\mathcal{L}(\mathcal{Q}_0) = \mathcal{L}(\mathcal{Q}_1)$, the following condition holds:
$$
\begin{aligned}
	\Big| \Pr[\mathcal{A}(\mathsf{View}_{\mathrm{P}}(\mathcal{Q}_0))=1] & 
	- \Pr[\mathcal{A}(\mathsf{View}_{\mathrm{P}}(\mathcal{Q}_1))=1] \Big| & \\ \leq \mathsf{negl}(\lambda).
\end{aligned}
$$
Here, $\mathsf{View}_{\mathrm{P}}(\mathcal{Q})$ contains the public execution plan, query ciphertexts, keyword matching traces, intermediate ciphertexts generated by slot routing and boolean evaluation, and response ciphertexts.

\noindent \textbf{Theorem 1.} 
Assuming the RLWE problem is hard, IBPIR satisfies IND-CKA security.

\textit{Proof.} We construct the following three hybrid experiments:
\begin{itemize}
	\item $Game_0$: Execute the real protocol on the keyword list $KL_0$ and output the real view $\mathsf{View}_{\mathrm{I}}(KL_0)$. 
	\item $Game_1$: Replace the query ciphertexts with ciphertexts obtained by encrypting zero-plaintext vectors of equal length, and generate the complete server view according to Algorithm \ref{alg:response}.
	\item $Game_2$: Replace the ciphertexts of zero plaintexts with the real query ciphertexts of $KL_1$. 
	The resulting distribution is the real server view $\mathsf{View}_{\mathrm{I}}(KL_1)$ of $KL_1$.
\end{itemize}

Since BFV is IND-CPA secure, the distinguishing advantages between $Game_0$ and $Game_1$, as well as between $Game_1$ and $Game_2$, are both negligible. 
Otherwise, the corresponding distinguisher could be directly used to distinguish BFV challenge ciphertexts. 
Therefore, the adversary's advantage in distinguishing $\mathsf{View}_{\mathrm{I}}(KL_0)$ and $\mathsf{View}_{\mathrm{I}}(KL_1)$ is at most some negligible function $\mathsf{negl}_{\mathrm{I}}(\lambda)$. 
Thus, IBPIR satisfies IND-CKA security. \hfill $\blacksquare$

\noindent \textbf{Theorem 2.} 
Assuming the RLWE problem is hard, PMKPIR satisfies IND-CBQA security.

\textit{Proof.} For any boolean query $\mathcal{Q}$, we construct the following real and ideal experiments.

\textbf{Real Experiment.} 
The client generates $\{\widetilde{Q}_\tau\}_{\tau=0}^{\alpha_{\mathcal{Q}}-1}$ according to the real protocol. 
The server sequentially executes keyword matching, slot routing, and boolean evaluation according to $\Pi_{\mathcal{Q}}$. 
This experiment outputs the real server view: $\mathsf{Real}_{\mathrm{P}}(\lambda,\mathcal{Q}) = \mathsf{View}_{\mathrm{P}}(\mathcal{Q}).$

\textbf{Ideal Experiment.} 
The simulator $\mathsf{Sim}$ only obtains the security parameter $\lambda$, public parameters $\mathsf{pp}$, encoded database $DB_{\mathrm{PMKPIR}}$, and leakage function $\mathcal{L}(\mathcal{Q})$. 
The simulator extracts $\Pi_{\mathcal{Q}}$ and $\alpha_{\mathcal{Q}}$ from the leakage function, and generates $\widetilde{Q}_\tau^{\,*} \leftarrow \mathsf{Encrypt}_{pk}(\mathbf{0}), 0 \leq \tau < \alpha_{\mathcal{Q}},$ where $\mathbf{0}$ is a BFV zero plaintext with the same slot dimension as the real query plaintext $Q_\tau$. 
Subsequently, the simulator executes the complete protocol taking $DB_{\mathrm{PMKPIR}}$, $\Pi_{\mathcal{Q}}$, and $\{\widetilde{Q}_\tau^{\,*}\}$ as input, and outputs the simulated server view: $\mathsf{Ideal}_{\mathrm{P},\mathsf{Sim}} (\lambda,\mathcal{L}(\mathcal{Q})) = \mathsf{Sim}(1^\lambda, \mathsf{pp}, DB_{\mathrm{PMKPIR}}, \mathcal{L}(\mathcal{Q})).$

Since BFV is IND-CPA secure, the real query ciphertexts and the ciphertexts generated by the simulator are computationally indistinguishable. 
Keyword matching, slot routing, and boolean evaluation are all public polynomial-time homomorphic operations, and their execution order is completely determined by $\Pi_{\mathcal{Q}}$. 
Therefore, for any PPT adversary $\mathcal{A}$, there exists a negligible function $\mathsf{negl}_{\mathrm{RI}}$ such that
$$
\begin{aligned}
	\Big| \Pr[\mathcal{A}(\mathsf{Real}_{\mathrm{P}}(\lambda,\mathcal{Q}))=1] & \\
	- \Pr[\mathcal{A}(\mathsf{Ideal}_{\mathrm{P},\mathsf{Sim}}(\lambda,\mathcal{L}(\mathcal{Q})))=1] \Big| &\leq \mathsf{negl}_{\mathrm{RI}}(\lambda).
\end{aligned}
$$

For $\mathcal{Q}_0$ and $\mathcal{Q}_1$ in Definition 2, identical leakage functions lead to identical distributions for the two ideal experiments. 
By the triangle inequality, we obtain:
$$
\begin{aligned}
	&\Big| \Pr[\mathcal{A}(\mathsf{View}_{\mathrm{P}}(\mathcal{Q}_0))=1] - \Pr[\mathcal{A}(\mathsf{View}_{\mathrm{P}}(\mathcal{Q}_1))=1] \Big| \\
	&\leq \Big| \Pr[\mathcal{A}(\mathsf{Real}_{\mathrm{P}}(\lambda,\mathcal{Q}_0))=1] \\
	&\quad\quad - \Pr[\mathcal{A}(\mathsf{Ideal}_{\mathrm{P},\mathsf{Sim}}(\lambda,\mathcal{L}(\mathcal{Q}_0)))=1] \Big| \\
	&\quad + \Big| \Pr[\mathcal{A}(\mathsf{Ideal}_{\mathrm{P},\mathsf{Sim}}(\lambda,\mathcal{L}(\mathcal{Q}_1)))=1] \\
	&\quad\quad - \Pr[\mathcal{A}(\mathsf{Real}_{\mathrm{P}}(\lambda,\mathcal{Q}_1))=1] \Big| \\
	&\leq 2\mathsf{negl}_{\mathrm{RI}}(\lambda) = \mathsf{negl}(\lambda).
\end{aligned}
$$

If there exists an adversary that can distinguish $\mathsf{View}_{\mathrm{P}}(\mathcal{Q}_0)$ and $\mathsf{View}_{\mathrm{P}}(\mathcal{Q}_1)$ with a non-negligible advantage, then there exists an adversary that can break the IND-CPA security of BFV with a non-negligible advantage, which contradicts the hardness assumption of RLWE. 
Therefore, PMKPIR satisfies IND-CBQA security. \hfill $\blacksquare$

\subsection{Response Privacy}

\noindent \textbf{Corollary 1.} 
Assuming the RLWE problem is hard, for any two keyword lists satisfying $\alpha_{KL_0}=\alpha_{KL_1}$, the response ciphertexts generated by IBPIR are computationally indistinguishable. 
Similarly, for any two boolean queries satisfying $\mathcal{L}(\mathcal{Q}_0)=\mathcal{L}(\mathcal{Q}_1)$, the response ciphertexts generated by PMKPIR are computationally indistinguishable. 

\textit{Proof.} 
The response ciphertexts are included in the server view. 
If two sets of response ciphertexts could be distinguished with a non-negligible advantage, then the server views containing these response ciphertexts could also be distinguished with the same advantage, which contradicts Theorem 1 and Theorem 2, respectively. 
Therefore, both IBPIR and PMKPIR satisfy the aforementioned security. \hfill $\blacksquare$

Furthermore, for an IBPIR query of length $s$, the server returns exactly $\alpha f_B$ response ciphertexts, where $\alpha=\lceil s/B\rceil$; for a PMKPIR boolean query, the server returns $f_B=\lceil m/w\rceil$ encrypted bitmap partitions. 
These quantities are determined solely by the number of queried keywords, the batching parameters, and the database scale, and are strictly independent of the specific keywords used in the query or the actual number of matching documents. 
Therefore, the server cannot deduce how many documents are included in the query result based on the number of response ciphertexts.

\subsection{Execution Plan Leakage}

The client sends the execution plan $\Pi_{\mathcal{Q}}$ along with the query ciphertexts to the server. 
The execution plan specifies the operations and their order for the server to evaluate the boolean query, including AND, OR, and NOT operations, the partitioning and padding of local DNFs, as well as the slot routing among operands. 
Therefore, the server can infer the logical structure of the query, the number of conditions, and the position of each condition in the computation from the execution plan, but it cannot deduce the specific keywords corresponding to these conditions. 
Furthermore, the server can determine $\alpha_{\mathcal{Q}}=\lceil s/B\rceil$ from the number of received query ciphertexts. 
For two queries satisfying $\mathcal{L}(\mathcal{Q}_0)=\mathcal{L}(\mathcal{Q}_1)$, the server executes identical computation steps and receives the same number of query ciphertexts. 
According to Theorem 2, the server cannot distinguish whether $\mathcal{Q}_0$ or $\mathcal{Q}_1$ is being executed based on the query ciphertexts, intermediate ciphertexts, or response ciphertexts. 
Consequently, PMKPIR has the same type of leakage as BKPIR: both reveal the logical structure of the query and the number of query ciphertexts, while concealing the specific queried keywords and the query results. 

\subsection{Security under Database Updates}

Let $DB_{\mathrm{PMKPIR}}^{(\eta)}$ denote the encoded database at the $\eta$-th update epoch. 
Following document insertions, deletions, or modifications to keyword-document relations, the server re-encodes all affected keyword plaintexts and bitmap plaintexts according to the $\textbf{Setup}$ phase. 
As long as the public keyword set $KW$, mapping $ID$, parameters $N, t, w, k, B$, and the database layout remain unchanged, both the client's $\textbf{QueryGen}$ algorithm and the server's online computation procedures require no modifications.

\noindent \textbf{Theorem 3.} 
When the public parameters and database layout remain unchanged, database updates do not affect the IND-CKA security of IBPIR and the IND-CBQA security of PMKPIR.

\textit{Proof.} 
Fix any update epoch $\eta$. 
The updated database $DB_{\mathrm{PMKPIR}}^{(\eta)}$ is held by the server and remains identical during the execution of the two challenge queries. 
The reductions in Theorem 1 and Theorem 2 can directly utilize $DB_{\mathrm{PMKPIR}}^{(\eta)}$ to generate the keyword matching traces, intermediate ciphertexts, and response ciphertexts. 
The only difference between the two challenge executions remains the query ciphertexts. 
Therefore, if the server could distinguish the two challenge queries after a database update, it could also distinguish the BFV challenge ciphertexts following the constructions in Theorem 1 or Theorem 2, which contradicts the IND-CPA security of BFV. 
Thus, database updates do not compromise the query privacy of IBPIR and PMKPIR. \hfill $\blacksquare$

\subsection{Mitigating Query-Structure Leakage}

As in BKPIR, PMKPIR treats the query size and boolean structure as public information. 
The number of query ciphertexts, $\alpha_{\mathcal Q}=\lceil s/B\rceil$, reveals the query size, while the public execution plan leaks the exact number of leaf nodes and the compiled boolean structure. 

To hide the number of query ciphertexts, the client can select a public upper bound $\bar{s}$, pad the query with dummy keywords, and always transmit $\bar{\alpha}=\left\lceil\bar{s}/B\right\rceil$ query ciphertexts. 
Since the dummy keywords are encrypted together with the actual queried keywords, the server cannot distinguish occupied logical sub-blocks from those used for padding. 
This prevents leakage of the actual number of query ciphertexts, but still cannot hide the boolean structure contained in $\Pi_Q$.

To achieve stronger structure privacy, the boolean query can be compiled into a fixed boolean circuit template. 
All templates contain the same number of leaf nodes and logic gates. 
Unused leaf nodes are filled with encrypted identity operands, while the logic gate types are represented by encrypted selector bits. 
Given input operands $x$ and $y$ and an encrypted selector bit $g$, AND and OR gates can be evaluated uniformly as $a=xy$, $o=x+y-xy$, and $z=a+g(o-a)$, where $g=0$ selects the AND gate and $g=1$ selects the OR gate. 
In this way, the server can observe only the public template capacity and cannot learn the actual query structure. 
This extension increases the number of ciphertext multiplications and the multiplicative depth and is therefore provided only as an optional strategy.

\end{document}